\documentclass[12pt]{article}

\usepackage{amsmath,epsfig,epsf,psfrag,natbib}
\usepackage{amssymb}
\usepackage[latin1]{inputenc}

\usepackage{amsbsy}
\usepackage{lscape}
\usepackage{color,latexsym,amsfonts,amssymb,amsthm,amscd,amsmath,graphicx}

\newcommand{\be}{\begin{equation}}
\newcommand{\ee}{\end{equation}}
\newtheorem{theorem}{Theorem}
\newtheorem{lemma}{Lemma}

\newtheorem{remark}{Remark}

\newtheorem{corollary}{Corollary}

\newcommand{\balpha}{\mbox{\boldmath $\alpha$}}
\newcommand{\cn}{\mbox{$\mathcal{N}$}}

\newcommand{\ba}{\begin{eqnarray}}
\newcommand{\ea}{\end{eqnarray}}
\newcommand{\bas}{\begin{eqnarray*}}
\newcommand{\eas}{\end{eqnarray*}}

\def\mathF{\mathcal{F}}
\def\mathM{\mathcal{M}}
\def\mathA{\mathcal{A}}
\def\mathB{\mathcal{B}}
\def\mathP{\mathcal{P}}
\def\mathS{\mathcal{S}}
\def\mathO{\mathcal{O}}
\def\mathG{\mathcal{G}}
\def\mathC{\mathcal{C}}
\def\mathS{\mathcal{S}}

\def\MM{\mathbb{M}}
\def\PP{\mathbb{P}}
\def\GG{\mathbb{G}}
\def\RR{\mathbb{R}}

\def\bbf{{\boldsymbol f}}

\def\pf{\qquad \textsl{Proof}:\ }
\def\endpf{\quad \blacksquare}
\def\epf{\quad $\blacksquare$}

\def\var{{\mbox{var}}}

\begin{document}

\title{\bf Maximum Smoothed Likelihood  Component Density Estimation in Mixture Models with Known Mixing Proportions}

\author{Tao Yu\footnote{Supported in part by NUS Grant R-155-000-134-112 and R-155-000-123-112.}, Pengfei Li\footnote{Supported in part by the
Natural Sciences and Engineering Research Council of Canada, and
by a startup grant from the University of Waterloo.},  and Jing Qin\\
National University of Singapore,  University of Waterloo, \\and National Institutes of Health
}

\date{\today}

\maketitle

\begin{abstract}
In this paper,
we propose a maximum smoothed likelihood method to estimate the component density functions
of mixture models,
in which the mixing proportions are known and may differ among observations.
The proposed estimates maximize a smoothed log likelihood function and  inherit all the important properties of probability density functions. A majorization-minimization algorithm is suggested to compute the proposed estimates numerically. In theory, we show that starting from any initial value,  this algorithm increases the smoothed likelihood function and further leads to estimates that maximize the smoothed likelihood function. This indicates the convergence of the algorithm. Furthermore, we theoretically establish the asymptotic convergence rate of our proposed estimators.
An adaptive procedure is suggested to choose the bandwidths in our estimation procedure. Simulation studies show that the proposed method is more efficient than the existing method in
terms of integrated squared errors. A real data example is further analyzed.

\end{abstract}


\noindent{\bf Running title:} MSL Component Density Estimation in Mixture Models.

\section{Introduction}

In this paper, we consider the data with the following mixture structure. Let $\{X_i,\balpha_i\}$, $i=1,\ldots,n$,  be independent and identically distributed (i.i.d.) copies of $\{X, \balpha\}$. For every $i=1,\ldots,n$, $X_i$ comes from one of the $M$ subpopulations with probability density functions (pdfs) $f_1(x),\ldots,f_M(x)$.
Denote by $\alpha_{i,j}$ the probability that $X_i$ is from the $j$th subpopulation and let $\balpha_i=(\alpha_{i,1},\ldots,\alpha_{i,M})^\tau$.
Clearly $\alpha_{i,j} \ge 0$ and $\sum_{j=1}^M \alpha_{i,j} = 1$.
In summary, the pdf of $X_i$ conditioning on $\balpha_i$ is given by
\begin{eqnarray}
X_i|\balpha_i \sim \sum_{j=1}^M\alpha_{i,j} f_j(x). \label{mix-model}
\end{eqnarray}
Practically, $\balpha_i$ is known, observable, or can be reliably estimated from other sources.
That is, conditioning on $\balpha_i$, $X_i$ follows a mixture model with known mixing proportions.
Our main interest in this paper is to estimate $f_1(x),\ldots,f_M(x)$ nonparametrically.

Recently, data of the mixture structure in (\ref{mix-model}) have been more and more frequently identified in the literature and in practice.
Acar and Sun (2013) provided one example of such data. In the genetic association study of single nucleotide polymorphisms (SNPs), the corresponding genotypes  of SNPs are usually not deterministic; in the resultant data, they are typically delivered as genotype probabilities from various genotype calling or imputation algorithms (see for example Li et~al. 2009 and Carvalho et~al. 2010).
Ma and Wang (2012) summarized two types of genetic epidemiology studies under which mixture data of such kind are collected. These studies are kin-cohort studies (Wang et~al.~2008) and quantitative trait locus studies (Lander and Botstein~1989; Wu et~al.~2007); see also Wang et~al.~(2012) and the references therein.
Section \ref{section-read-data} also gives an example of such data in the malaria study. More examples and the corresponding statistical research can be founded in Ma et~al. (2011), Qin et~al. (2014), and the references therein.

With the data of the mixture structure in (\ref{mix-model}), statistical methods for estimating the component cumulative distribution functions (cdfs) have been established in the literature. A comprehensive overview of these developments is as follows.
Ma and Wang (2012) pointed out that the classic maximum empirical  likelihood estimators of these component cdfs
are either highly inefficient or inconsistent.
They proposed a class of weighted least square estimators.
Wang et al. (2012) and Ma and Wang (2014)
proposed consistent and efficient
nonparametric estimators based on estimating equations
for the  component cdfs
when the data are censored.
Qin et~al. (2014) considered another class of estimators for the component cdfs by maximizing the
binomial likelihood.
Their method can be applied to data with censored or a non-censored structure.
We observe that all these works were focused on the estimation of cdfs and assumed $\balpha_i$ to be a discrete random vector.

The estimation of the pdfs are less addressed in the literature. As far as we are aware, to date Ma et~al.~(2011) is the only existing reference that considered the component density estimation under the setup of model (\ref{mix-model}).
They proposed a family of kernel-based weighted least squares
estimators for the component pdfs under the assumption that $\balpha_i$ is continuous.
However, to the best of our knowledge, there are two limitations in their approach: (1) the resultant estimates do not inherit the nonnegativity property of a regular density function; as is well known, such a property is often important in many downstream density-based studies.
In that paper, though authors have discussed an EM-like algorithm to achieve nonnegative component density estimates, 
the corresponding theoretical properties as well as the numerical performance of these estimates were not studied yet. (2) When dealing with some practical problems, this method does not make full use of the data and therefore the resultant density estimation may not be as efficient. We refer to the end of Section \ref{section-sim} for an example and further discussion.

In this paper, we consider maximum smoothed likelihood (Eggermont and Lariccia 2001, Chapter 4) estimators for $f_1, \ldots, f_M$, namely $\widehat f_1, \ldots, \widehat f_M$, which maximize a smoothed likelihood function and  inherit all the important properties of pdfs.
Our method can handle data with $\balpha_i$'s continuous or discrete.
We also propose a majorization-minimization algorithm that computes these density estimates numerically. This algorithm incorporates similar ideas as Levine et~al.~(2011) and the EM-like algorithm (Hall et~al.~2005). 
We
show
that under finite samples, starting from any initial value,  this algorithm not only increases the smoothed likelihood function but also leads to estimates that maximize the smoothed likelihood function.

Another main contribution of this paper is to establish the $L_1$ asymptotic consistency and the corresponding convergence rate for our density estimates. Because of the properties (see Section \ref{section-3}) of the non-linear operator ``$\cn_h$" defined in Section \ref{section-2} and the complicated form of the smoothed log-likelihood function, the development of the asymptotic theories for the nonparametric density estimates under the framework of mixture model is technically challenging and still lacking in the literature. We solve this problem by 
employing the advanced theories in empirical process (see van der Vaart and Wellner~1996, Kosorok~2008, and the references therein). We expect that the technical tools established in this paper may benefit the future study on the asymptotic theories of the nonparametric density estimates for mixture model of other kinds; see for example Levine et~al.~(2011).

The rest of the paper is organized as follows.  Section \ref{section-2} presents our proposed density estimates based on smoothed likelihood principal. Section \ref{section-mm-alg} suggests a majorization-minimization algorithm to numerically compute these density estimates, and establishes the finite-sample convergence properties of this algorithm. Section \ref{section-3} studies the asymptotic behaviors of our density estimators. Section \ref{section-band-sel} proposes a bandwidth selection procedure that is easily imbedded into the majorization-minimization algorithm. Section \ref{section-sim} conducts simulation studies, which show that the proposed method is more efficient than existing methods in terms of integrated square error. Section \ref{section-read-data} applies our method to a real data example.
The technical details are relegated to the Appendix.

%

\section{Maximum Smoothed Likelihood Estimation} \label{section-2}

With the observed data $\{X_i, \balpha_i\}_{i=1}^n$ from Model (\ref{mix-model}),  we propose a maximum smoothed likelihood method for estimating $f_1, \ldots, f_M$.
We consider the set of functions
\begin{eqnarray*}
\mathC = \left\{(f_1,\ldots, f_M): f_j\mbox{ is a pdf  },j=1,\ldots,M \right\}.
\end{eqnarray*}
Furthermore, we assume that $f_j$'s have the common support $S_x$.

Given Model (\ref{mix-model}) and observations  $\{X_i,\balpha_i\}_{i=1}^n$, the conditional log-likelihood can be written as
\begin{eqnarray*}
\widetilde l_n(f_1, \ldots, f_M) = \sum_{i=1}^n \log\left\{\ \sum_{j=1}^M \alpha_{i,j} f_j(X_i) \right\}.
\end{eqnarray*}
However, as is well known, this log-likelihood function is unbounded in $\mathC$; see page 25 in Silverman (1986) and  page 111 in Eggermont and Lariccia (2001).
Therefore, the corresponding maximum likelihood estimates do not exist.
This unboundedness problem can be solved by incorporating the smoothed likelihood approach (Eggermont and Lariccia, 1995, Groeneboom et~al.~ 2010, Yu et al.~2014, and the references therein). Specifically, we define the smoothed log-likelihood of $f_1(x),\ldots, f_M(x)$ to be
\begin{equation}
l_n(f_1,\ldots,f_M)=\sum_{i=1}^n\log \left\{\sum_{j=1}^M\alpha_{i,j}\cn_{h_j} f_j(X_i) \right\}, \label{def-l-n}
\end{equation}
where $\cn_h f(x)$ is the nonlinear smoothing operator for a density function $f$, represented by
\begin{eqnarray}
\cn_h f(x)&=&\exp\left\{\int_{\RR} K_h(u-x)\log f(u) du\right\}. \label{def-cn}
\end{eqnarray}
Here $K_h(x)=\frac{1}{h}K(x/h)$, $K(\cdot)$ is a kernel function supported on $[-L, L]$, and $h$ is the bandwidth for the nonlinear smoothing operator. By convention, we define $0\log(0)=0$, $\log(0)=-\infty$, and $\exp(-\infty)=0$.

Our proposed maximum smoothed likelihood estimators for $f_1,\ldots,f_M$ are given by
\begin{eqnarray}
(\widehat f_1,\ldots,\widehat f_M)
=\mbox{argmax}_{(f_1,\ldots,f_M) \in \mathC} l_n(f_1,\ldots,f_M). \label{def-f-hat}
\end{eqnarray}
We observe that  the smoothed likelihood function defined in (\ref{def-l-n}) has the following properties. First, based on Lemma 3.1 (iii) of Eggermont (1999), $l_n(\cdot)$ is concave in $\mathC$, and $\mathC$ is a convex set of functions.
Second, if the kernel function $K(t)$ is bounded and $h_j>0$, $j=1,\ldots,M$ are fixed, then $l_n(\cdot)$ is also bounded in $\mathC$, since for every $x$ and $(f_1, \ldots, f_M)\in \mathC$,
\begin{eqnarray*}
\cn_{h_j}f_j(x)
&\le& \exp\left[ \log \left\{\int_{\RR} K_h(u-x) f(u)du\right\} \right]\le  \sup_t K(t)/h_j.
\end{eqnarray*}
%
Therefore, the maximizer of $l_n(\cdot)$ exists, i.e., the optimization problem (\ref{def-f-hat}) is well defined. Furthermore, if we assume that for every $j=1,\ldots,M$, the $X_i$'s corresponding to $\alpha_{i,j} > 0$ are dense in $S_x$,
then $l_n(\cdot)$ is strictly concave in $\mathC$ and thus the solution to the optimization problem (\ref{def-f-hat}) is unique. Here, ``dense" means for every $j=1,\ldots,M$, and $x\in S_x$, the interval $[x-Lh_j, x+Lh_j]$ contains at least one observation $X_i$, such that the corresponding $\alpha_{i,j}>0$.

\section{The majorization-minimization algorithm} \label{section-mm-alg}
In this section, we propose an algorithm that numerically calculates $\widehat f_1,\ldots,\widehat f_M$
with given bandwidths $h_1,\ldots,h_M$ and study the finite-sample convergence property of this algorithm.
The proposed algorithm, called the majorization-minimization algorithm,  is in spirit similar to the majorization-minimization algorithm in Levine~et~al. (2011).
 To facilitate our theoretical development, we define the majorization-minimization updating operator $\mathG$ on $\mathC$ as follows. For any $(f_1, \ldots, f_M) \in \mathC$, let
 \begin{eqnarray}
 \mathG(f_1,\ldots, f_M) = (f_1^{\mathG}, \ldots, f_M^{\mathG}), \label{def-mathG}
\end{eqnarray}
where
\begin{eqnarray}
\nonumber
f_{j}^{\mathG} (x)&=&\frac{\sum_{i=1}^n w_{i,j}K_{h_j}(x-X_i)}{\sum_{i=1}^n w_{i,j}}, \label{minization} \\
\mbox{with }\quad w_{i,j}&=&\frac{\alpha_{i,j}\cn_{h_j}f_j(X_i)}{\sum_{k=1}^M\alpha_{i,k}\cn_{h_k} f_k(X_i)}.
\label{wij.post}
\end{eqnarray}
We first show that $\mathG$ is capable of increasing the smoothed log-likelihood function $l_n$ in every step of updating.
\begin{theorem} \label{theorem-1}
For every $(f_1,\ldots, f_M) \in \mathC$, we have $l_n\left(\mathG(f_1,\ldots,f_M)\right)\geq l_n\left(f_1,\ldots,f_M\right)$.
\end{theorem}
Theorem \ref{theorem-1} immediately leads to our proposed majorization-minimization algorithm as follows.
Given initial values $(f_1^{0},\ldots,f_M^{0}) \in \mathC$, for $s=0,1,2,\cdots$, we iteratively update from $(f_1^{s}, \ldots, f_M^{s})$ to $(f_1^{s+1}, \ldots, f_M^{s+1})$ as $$(f_1^{s+1}, \ldots, f_M^{s+1}) = \mathG(f_1^{s}, \ldots, f_M^{s}).$$
Clearly, Theorem \ref{theorem-1} above ensures that for every $s=0, 1, \ldots$, we have $l_n\left(f_1^{s+1},\ldots,f_M^{s+1}\right)\geq l_n\left(f_1^{s},\ldots,f_M^{s}\right)$. Furthermore, since for any $(f_1,\ldots, f_M) \in \mathC$, $\mathG(f_1,\ldots, f_M)$ belongs to the class of functions $\mathF_n$:
\begin{eqnarray}\mathF_n &=& \left\{(f_1, \ldots, f_M): f_j(x) = \frac{\sum_{i=1}^nw_{i,j} K_{h_j}(x-X_i)}{\sum_{i=1}^n w_{i,j}}; 0 \le w_{i,j} \le 1\right\}, \label{def-mathF-n}
\end{eqnarray}
therefore, $(f_1^{s}, \ldots, f_M^{s})\in \mathF_n$ for $s\ge 1$. Next, we study the finite-sample convergence property of this majorization-minimization algorithm; we observe that the technical development for such a convergence property is nontrivial. We first present a sufficient and necessary condition under which $(\widehat f_1,\ldots,\widehat f_M)\in \mathC$ is a solution of the optimization problem (\ref{def-f-hat}).

\begin{theorem} \label{theorem-1-added-1}
Assume for every $j=1,\ldots, M$, $\sum_{i=1}^n \alpha_{i,j} >0$. Consider $(\widehat f_1, \ldots, \widehat f_M)\in \mathC$, then
\begin{eqnarray*}
l_n(\widehat f_1,\ldots,\widehat f_M)
=\sup_{(f_1,\ldots,f_M) \in \mathC} l_n(f_1,\ldots,f_M)
 \end{eqnarray*}
 if and only if $(\widehat f_1, \ldots, \widehat f_M) = \mathG(\widehat f_1, \ldots, \widehat f_M)$ almost surely under the Lebesgue measure.
\end{theorem}
The following corollary is resulted from an immediately application of Theorem \ref{theorem-1-added-1}; the straightforward proof is omitted.
\begin{corollary} \label{corollary-1}
Assume for every $j=1,\ldots, M$, $\sum_{i=1}^n \alpha_{i,j} >0$. Let $(\widehat f_1, \ldots, \widehat f_M)$ be a solution of the optimization problem (\ref{def-f-hat}), then $(\widehat f_1, \ldots, \widehat f_M)\in \mathF_n$ almost surely under the Lebesgue measure.
\end{corollary}

Corollary \ref{corollary-1} benefits our subsequent technical development of the asymptotic theories for $\widehat f_1, \ldots, \widehat f_M$ in Section \ref{section-3}. It indicates that the solution of (\ref{def-f-hat}) is equivalent to the solution of
\begin{eqnarray}
(\widehat f_1,\ldots,\widehat f_M)
=\mbox{argmax}_{(f_1,\ldots,f_M) \in \mathF_n} l_n(f_1,\ldots,f_M), \label{def-f-hat-1}
\end{eqnarray}
as long as the stated condition $\sum_{i=1}^n \alpha_{i,j} >0$ for every $j$  is satisfied.
This condition is quite reasonable since  if $\sum_{i=1}^n \alpha_{i,j} =0$ for some $j$
then the $j$th subpopulation does not appear in the data and we can delete the corresponding $f_j(x)$ from the mixture model (\ref{mix-model}).
%
Therefore, developing the asymptotic theories for $\widehat f_1, \ldots, \widehat f_M$ from (\ref{def-f-hat}) is equivalent to developing those from (\ref{def-f-hat-1}).

Based on Theorem \ref{theorem-1-added-1}, we show the convergence of the updating sequence $l_n\left(f_1^{s},\ldots,f_M^{s}\right)$ to its global maximum, which implies the convergence of
the proposed majorization-minimization algorithm.
\begin{theorem} \label{theorem-1-added-2}
Assume $\sup_t K(t) < \infty$. Then, we have
\begin{eqnarray*}
\lim_{s\to \infty}l_n(f_1^{s}, \ldots, f_M^{s}) =  l_n(\widehat f_1, \ldots, \widehat f_M), \label{thm-1-added-2-0}
\end{eqnarray*}
where $(\widehat f_1, \ldots, \widehat f_M)\in \mathF_n$ is a solution of the optimization problem (\ref{def-f-hat}).
\end{theorem}

Based on Theorem \ref{theorem-1-added-2}, if we don't impose further conditions on the data, $l_n(\cdot)$ is not necessarily strictly concave. Therefore, we can only show that the updating sequence $l_n(f_1^{s}, \ldots, f_M^{s})$ converges to the maximum of $l_n(f_1, \ldots, f_M)$. Note that this does not guarantee the convergence of $(f_1^{s}, \ldots, f_M^{s})$ to $(\widehat f_1, \ldots, \widehat f_M)$, i.e., the maximizer of $l_n(f_1, \ldots, f_M)$, because  such a maximizer may not be uniquely defined.
Instead, referring to the proof of this theorem, we have shown that there exists at least a subsequence of $(f_1^{s}, \ldots, f_M^{s})$ converging to a maximizer of $l_n(f_1, \ldots, f_M)$.
Furthermore, if we impose some technical condition such that $l_n(\cdot)$ is strictly concave, $(\widehat f_1, \ldots, \widehat f_M)$ is then uniquely defined by (\ref{def-f-hat}). Immediately, we can show $\lim_{s\to \infty}(f_1^{s}, \ldots, f_M^{s}) = (\widehat f_1, \ldots, \widehat f_M)$ for every $x\in S_x$. We refer to the discussion at the end of Section 2 for a sufficient condition so that $l_n$ is strictly concave in $\mathC$.

We end this section with the following remark about the proposed majorization-minimization algorithm above.
\begin{remark}
Ma et al. (2011) discussed an EM-like algorithm in their discussion section
to obtain nonnegative component density estimates.
In particular, they suggested defining
$$
w_{i,j}=\frac{\alpha_{i,j}f_j(X_i)}{\sum_{k=1}^M\alpha_{i,k}f_k(X_i)},
$$
and using a similar way as (\ref{def-mathG}) to update the resultant density estimates in their paper.
Yet, the corresponding theoretical properties as well as the numerical performance of these estimates are left unknown.
As commented by Levine et al. (2011), algorithms of this kind do not minimize/maximize any particular
objective function; this may impose difficulty in the subsequent technical development.
We refer to Levine et al. (2011) for more discussion of such a method.
\end{remark}

\section{Asymptotic Properties for $(\widehat f_1,\ldots,\widehat f_M)$}\label{section-3}

In this section, we investigate the asymptotic behaviors of $(\widehat f_1,\ldots,\widehat f_M)$ given in (\ref{def-f-hat}).
First, we consider the consistency of $\widehat p(x, \balpha)=\sum_{j=1}^M\alpha_j \cn_{h_j} \widehat f_j(x)$
under the Hellinger distance,
where the Hellinger distance between two non-negative functions $g_1(\omega)$ and $g_2(\omega)$ is defined to be
\begin{eqnarray*}
d(g_1, g_2) = \left[\int_{\Omega} \left\{g_1^{1/2}(\omega) - g_2^{1/2}(\omega) \right\}^2 d\mu\right]^{1/2}.
\end{eqnarray*}
where $g_1, g_2$ are functions defined on $\Omega$, $\mu$ is a measure on $\Omega$.

\begin{theorem} \label{consistency-theorem-1}
Assume Conditions 1--3. Then for any $\vartheta>0$, we have
\begin{eqnarray*}
d(\gamma\widehat p, \gamma \widetilde p_0) = O_p(h^{0.5}) + O_p(n^{-0.5+\vartheta}h^{-0.5}).
\end{eqnarray*}
where $\gamma(\balpha)$ is the marginal density of $\balpha$,
$\widetilde p_0(x, \balpha) = \sum_{j=1}^M \alpha_j f_{0,j}(x)$ is the conditional density of $X$ given $\balpha$,  and $f_{0,j}(x)$, $j=1,\ldots,M$, denote the true values of $f_j(x)$.
\end{theorem}


Next we establish the asymptotic convergence rate for $\cn_{h_j} \widehat f_j$, $j=1,\ldots,M$ under the $L_1$-distance. The proof of this theorem heavily replies on the results given in Theorem \ref{consistency-theorem-1}.

\begin{theorem}\label{theorem-3}
Assume Conditions 1--4 in Appendix B. For every $\vartheta > 0$ and $j=1,\ldots,M$, we have
\begin{eqnarray*}
\int_{\RR} |\cn_{h_j} \widehat f_j(x)- f_{0,j}(x)|dx = O_p(h^{1/2}) + O_p(n^{-0.5 + \vartheta}h^{-0.5}).
\end{eqnarray*}

\end{theorem}
Last, we establish the  $L_1$ convergence of $\widehat f_j(x)$. We observe that the results by Theorems \ref{theorem-1-added-1} and \ref{theorem-3} play key roles in the proof.
\begin{theorem} \label{theorem-4}
Assume Conditions 1--4 in Appendix B. For any $\vartheta > 0$, we have
\begin{eqnarray*}
\int_{\RR} |\widehat f_j(x)- f_{0,j}(x)|dx = O_p(h^{1/2}) + O_p(n^{-0.5 + \vartheta}h^{-0.5}),~~j=1,\ldots,M.
\end{eqnarray*}
\end{theorem}

For presentational continuity, we have organized the technical conditions and long proofs of Theorems \ref{consistency-theorem-1}--\ref{theorem-4} in Appendix B.  As observed in Appendix B, the theoretical developments for these theorems are technically challenging. The main obstacles are due to the following undesirable properties of $\cn_h f(x)$ with $f(x)$ being an arbitrary pdf.

Firstly, $\cn_{h}f(x)$ is neither a density nor necessarily sufficiently close to the corresponding $f(x)$. Therefore,  the well developed empirical process theories and technics for M-estimators in density estimation (see for example Section 3.4.1 in van der Vaart and Wellner 1996) is not directly applicable.

Secondly, $\cn_h f(x)$ introduces significant bias on the boundary of the support of $f(x)$. For example, if $f(x)$ is supported on $[c_1, c_2]$, then $\cn_h f(x)$ is supported on $[c_1+Lh, c_2-Lh]$. That is $\cn_hf(x) = 0$ when $x\in [c_1, c_1+Lh)\cup(c_2-Lh, c_2]$.  Here $[-L, L]$ is the support for the kernel function $K(x)$.

These two properties of $\cn_h f(x)$ significantly challenge our technical development. So far, we can only show the asymptotic behaviours of $\widehat p(x)$, $\cn_{h_j} \widehat f_j(x)$, and $\widehat f_j(x)$ as those given in Theorems \ref{consistency-theorem-1}, \ref{theorem-3} and \ref{theorem-4}. The convergence rate given in Theorems \ref{theorem-3} and \ref{theorem-4} may not be the optimal. There is some room to improve. However,
because of these two properties of ``$\cn_h$", we conjecture that $O_p(h^{0.5})$ is the best rate achievable by $d(\gamma \widehat p, \gamma \widetilde p_0)$ under the assumption that $f_{0,j}(x)$'s are supported on a compact support.  The intuition is as follows. Consider the extreme case that even though $\widehat f_j(x)$'s are estimated ideally well, $\widehat f_j(x) = f_{0,j}(x)$ say, one can show that the best rate for $d(\gamma \widehat p, \gamma \widetilde p_0)$ is still bounded by $O_p(h^{0.5})$.

\section{Bandwidth Selection} \label{section-band-sel}
The maximum smoothed likelihood estimates $\widehat f_1,\ldots,\widehat f_M$ depend on the choice of the bandwidths $h_1,\ldots,h_M$.
We suggest an algorithm that embeds the selection of the bandwidth $h_1,\ldots,h_M$ into the updating steps of the majorization-minimization algorithm suggested in Section \ref{section-mm-alg}.

Let $n_j$ be the positive integer closest to $\sum_{i=1}^n\alpha_{i,j}$,
which serves as an estimate of the average number of observations from the $j$th population.
Given initial values $(f_1^{0},\ldots,f_M^0)$ and initial bandwidths $(h_1^0, \ldots, h_M^{0})$, for $s=0,1,2,\cdots$. We update $(f_1^{s},\ldots,f_M^{s})$ and $(h_1^{s}, \ldots, h_M^{s})$ as follows.

\begin{enumerate}
\item[] Step 1.
For every $i = 1,\ldots,n$ and $j = 1,\ldots,M$, let
$$
w_{i,j}^{s}=\frac{\alpha_{ij}\cn_{h_j^{s}}f_j^{s}(X_i)}{\sum_{j=1}^M\alpha_{ij}\cn_{h_j^{s}} f_j^{s}(X_i)}.
$$

\item[] Step 2.
Sort $w_{i,j}^{s}$:
$w_{(1),j}^{(s)}\geq w_{(2),j}^{s}\geq\cdots \geq w_{(n), j}^{s}$.
Let $S_j^{s}=\{X_i: w_{i,j}\geq w_{(n_j), j}^{s}\}$. Treating the observations in $S_j^{s}$ as from a single population,
 we apply available bandwidth selection method for classical kernel density estimate to choose $h_j$.
Denote by $h_j^{s+1}$ the resultant   bandwidth.

\item[] Step 3. Let
$$
f_{j}^{s+1} (x)=\frac{\sum_{i=1}^n w_{ij}^{s} K_{h_j^{s+1}}(x-X_i)}{\sum_{i=1}^n w_{ij}^{s}}.
$$
\end{enumerate}
The philosophy of the above bandwidth selection step (i.e. Step 2) is as follows. In fact, $S_j^s$ collects the $n_j$ observations most likely coming from the $j$th population based on the preceding iteration. Therefore, we use these observations to select the bandwidth for the corresponding density estimates in the current iteration.

When implementing this algorithm in our numerical studies,
we use the quartic kernel, which was also used by Ma et al. (2011).
The initial density $(f_1^{0},\ldots,f_M^0)$ is randomly chosen from $\mathF_n$, i.e.,
the corresponding weights $w_{i,j}$ are randomly generated from the uniform distribution over [0,1]. In the bandwidth selection step (i.e. Step 2), once $S_j^s$ is obtained, we use R function {\tt dpik()}
to update the bandwidths $h_j^{s+1}$, $j=1,\ldots, M$.
{\tt dpik()}  in the R package {\tt KernSmooth} is implemented by Wand and Matt (publicly available at {\tt http://CRAN.R-project.org/package=KernSmooth}). This package is based on the kernel methods in Wand and Jones (1996).
Furthermore, the initial bandwidths are set as $h_j^0 = h^0$ for every $j=1,\ldots, M$, where $h^0$ is the output of {\tt dpik()}
based on all the observations $X_1,\ldots,X_n$.
We iterate Steps 1--3 until the change of the smoothed likelihood $l_n(f_1,\ldots,f_M)$  is smaller than a tolerance value $10^{-5}$ in each iteration.

In our numerical studies,  we observe that this algorithm converges fast. For example, consider the real data example in Section \ref{section-read-data}.
Setting the random seed set as ``123456",
 the bandwidths do not change up to 6th decimal point in two iterations;
the change of $l_n$ is less than $10^{-5}$ in another 59 iterations.
We have also experimented with other random seeds.
The results are very similar.
In addition, the resultant estimates for $f_1,\ldots, f_M$ are independent of the choice of  $(f_1^{0},\ldots,f_M^0)$.

\section{Simulation Study} \label{section-sim}
We use the following simulation examples to examine the numerical performance of our density estimates.
We consider three ``Studies".
Studies I and II adopt the same setup as those in Ma et~al. (2011)
so that we can compare the results by our method with those in that paper.
Study III mimics the real data example given in Section \ref{section-read-data}.

In the first study (\underline{Study~I}), we generate data using two populations, i.e., $M=2$.
Both populations have a standard normal distribution, so that $f_{0,1} = f_{0,2} = \phi_0$, where $\phi_0$ denotes the pdf of the standard normal distribution. We generate $X_1,\ldots, X_n$ with $n=400$. For every $X_i$, we set $\balpha_{i} = (\alpha_{i,1}, \alpha_{i,2})^\tau$ with $\alpha_{i,1}= u_{i,1}/(u_{i,1}+ u_{i,2})$, where $u_{i,1},  u_{i,2}$ are generated independently
from the uniform distribution over $[0,1]$. Therefore, approximately $200$ observations will come from each of the population. We repeat the simulation 1000 times and therefore obtain 1000 replicated simulation data sets $\{X_i, \balpha_i\}_{i=1}^{400}$.

For the second study (\underline{Study~II}), the settings are the same as those in Study I except that the two populations are simulated very differently. In particular, the distribution for the first population is simulated as normal distribution with mean 10 and variance 25, whereas the second is as a $t$ distribution centered at 20 with degrees of freedom 4 and scale parameter $10$.

\begin{figure}[!http]
\caption{
Pointwise quantile density estimates for Studies I (top panels), II (middle panels), and III (bottom panels).
In each plot, the solid line is the true density and the other three curves
are the pointwise quantiles for density estimates over 1000 replicates: median (dotted), 5\% (dashed), and 95\% (dashed).
}
\vspace{-0.5in}
\centerline{ \includegraphics[scale=0.7,angle=-90]{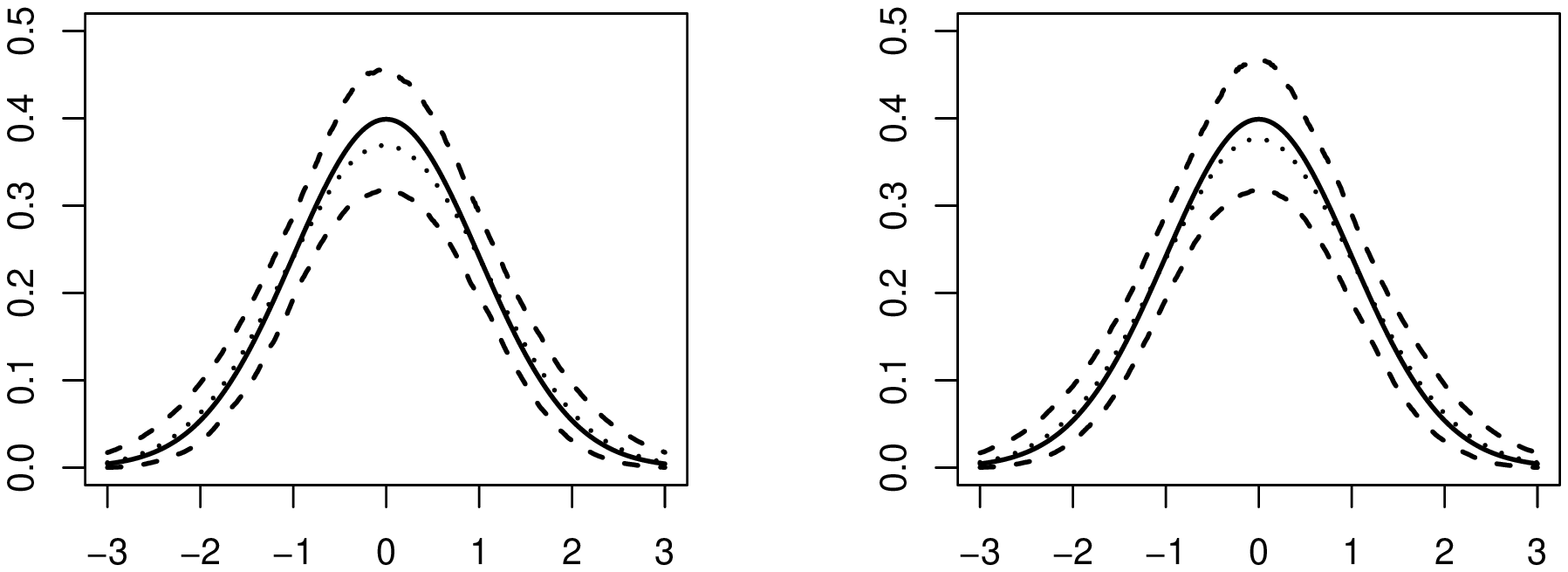}}
\vspace{-0.5in}
\centerline{ \includegraphics[scale=0.7,angle=-90]{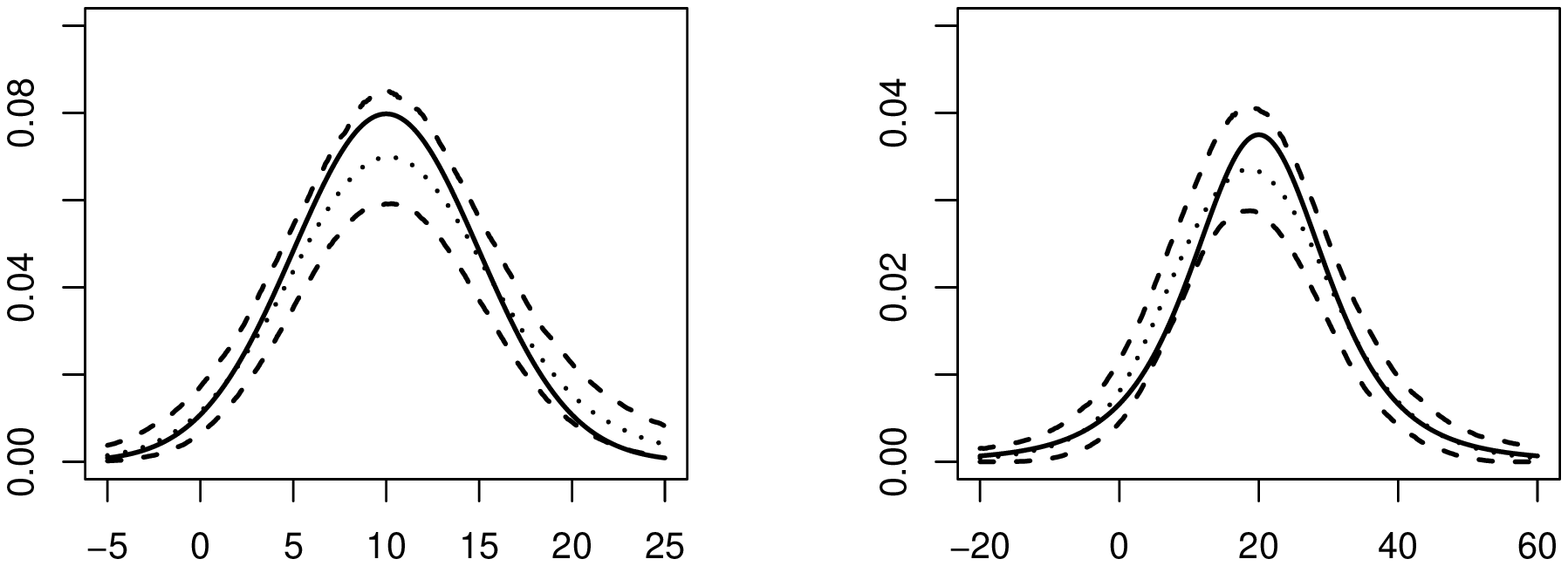}}
\vspace{-0.5in}
\centerline{ \includegraphics[scale=0.7,angle=-90]{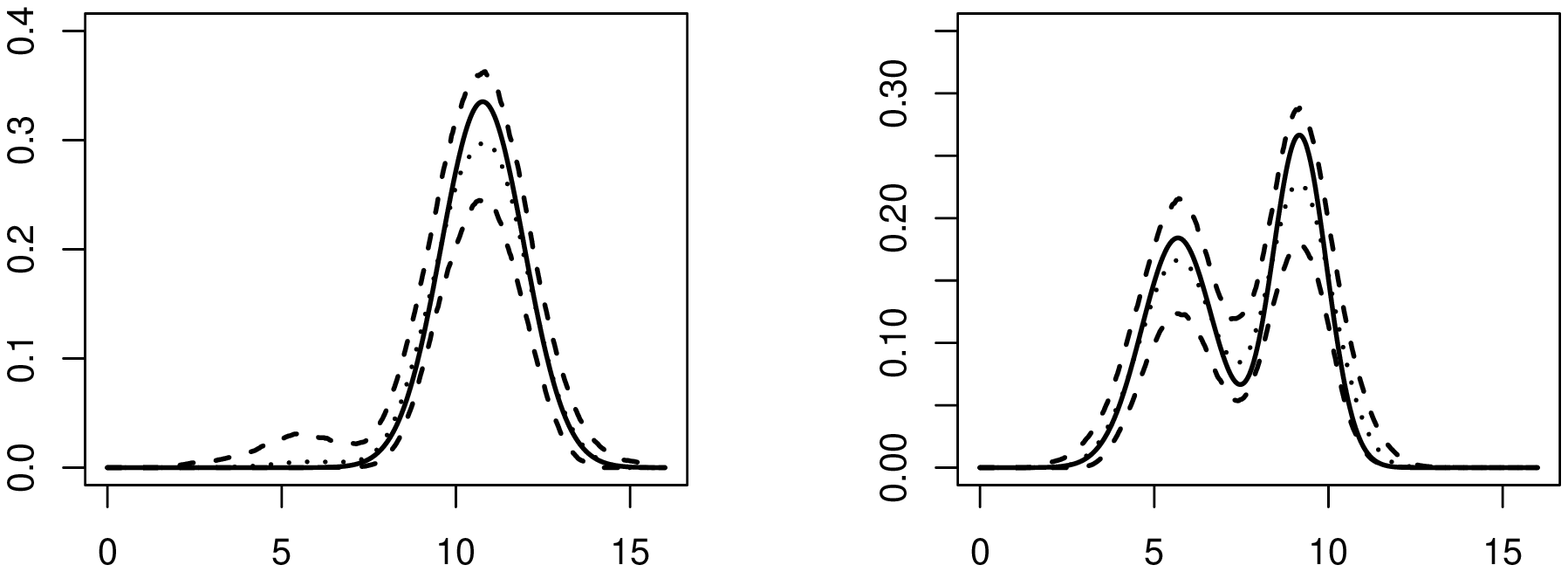}}
\label{pos}
\end{figure}

For every simulated data above, we apply the algorithm in Section \ref{section-band-sel} to obtain $\widehat f_1$ and $\widehat f_2$. The 5\%, 50\%, and 95\% point-wise quantiles for $\widehat f_1$ (left panel) and $\widehat f_2$ (right panel) over 1000 replications are given in Figure \ref{pos} (Study~I: top panels; Study~II: middle panels).
We notice that the 90\% confidence bands of $f_1$ and $f_2$ cover the corresponding true density.
To compare with the methods proposed by Ma et~al. (2011), we compute the average values  of integrated squared error (ISE) for $\widehat f_1$ and $\widehat f_2$ over 1000 replications in Studies I and II.
The results together with those presented by Ma et~al. (2011) are displayed in Table \ref{table-1}. For presentational brevity, in Table \ref{table-1}, we have only listed two proposed methods (namely ``OLS, ICV" and ``OLS, plug-in") in that paper, since for other methods the displayed results are very similar or even worse than these. Here ISE is defined to be
\begin{eqnarray*}
ISE(\widehat f_j) = \int \{\widehat f_j(x)- f_{0,j}(x)\}^2 dx.
\end{eqnarray*}
Overview Table \ref{table-1}, we have clearly observed that for Studies I and II, our method leads to smaller ISE values than methods proposed by Ma et~al.~(2011). The improvement is significant, particularly for the case that $f_{0,1}$ and $f_{0,2}$ are simulated similarly (i.e., Study~I).

\begin{table}[!http]
\caption{Average values of the integrated square error (ISE) for the proposed method  and Ma et al. (2011)'s method
in combination with ICV and adapted plug-in method under Studies I and II. Each value
in the table was computed from 1000 replications and is 100 $\times$ the actual value.} \label{table-1}
\begin{center}
{\small
\begin{tabular} {r|cc|cc}
\hline
& \multicolumn{2}{c|}{Study~I, $n=400$}& \multicolumn{2}{c}{Study~II, $n=400$} \\
Methods               & $f_1$&$f_2$&$f_1$&$f_2$\\
\hline
Ma et al. (2011): OLS, ICV& 0.73&0.73&0.19&0.07\\
Ma et al. (2011): OLS, plug-in&0.82&0.83&0.21&0.08\\
Our method&0.52&0.51&0.15&0.07\\
\hline
\end{tabular}
}
\end{center}
\end{table}

In the third study (\underline{Study~III}), we simulate densities that mimic the shape of those estimated from the real data example in Section \ref{section-read-data}.
The data are generated by: \begin{eqnarray*}
X_i|\alpha_i \sim f_2(x) \quad &\mbox{when }& i>n_1\\
X_i|\alpha_i \sim 0.677 f_1(x)+0.323f_2(x) \quad &\mbox{when }& i\le n_1,
\end{eqnarray*}
where $n_1=211$, $n_2=81$, $f_1(x)$, and $f_2(x)$ are the pdfs of $N(10.77,1.19)$ and
$0.48N(5.68,1.04)+0.52N(9.17,0.78)$ respectively.
Here, $N(\mu,\sigma)$ denotes normal distribution with mean $\mu$ and variance $\sigma^2$.
We choose these $f_1(x)$ and $f_2(x)$ so that they have similar shapes as those estimated from the real data example in Section \ref{section-read-data}.
We repeat the simulation 1000 times and therefore obtain 1000 replicated simulation data sets $\{X_i, \balpha_i\}_{i=1}^{292}$.

For each simulated data presented above,
a simple method to estimate $f_1(x)$ and $f_2(x)$ is as follows. Let $\widetilde f_2(x)$ be the kernel density estimate of $f_2(x)$ based on
$X_{n_1+1},\ldots, X_{n}$
and $\widetilde r(x)$ be that of $0.677 f_1(x)+0.323f_2(x)$ based on $X_{1},\ldots, X_{n_1}$.
We then estimate $f_1(x)$ by
$$
\widetilde f_1(x)=\frac{\widetilde r(x)-0.323\widetilde f_2(x)}{0.677}.
$$
This method is introduced in the introduction section of Ma et al. (2011).
Obviously, $\widetilde f_1(x)$ is not necessarily to be nonnegative.

We compare the results by our method with those by the simple method above.
When implementing the simple method, we use the quartic kernel and R function {\tt dpik()}
to obtain the bandwidths for $\widetilde f_2(x)$ and $\widetilde r(x)$.
The average ISEs over 1000 replications for $\widehat f_1$ and $\widehat f_2$
are both about $0.66\times 10^{-2}$. In contrast,
those for $\widetilde  f_1$ and $\widetilde f_2$
are $0.68\times 10^{-2}$ and $0.87\times 10^{-2}$ respectively. These observations are not surprising.
Appropriately accounting for the information  carried by $X_1,\ldots,X_{n_1}$, which is not used by $\widetilde f_2$, our method decreases the ISE of the estimate of $f_2(x)$ by about 24\%.
In contrast, $\widetilde f_1$ has fully used the information carried in $X_{n_1+1}, \ldots, X_n$; without extra information on $f_1(x)$, our method does not significantly outperform the simple method.
However, $\widehat f_1(x)$ is guaranteed to be nonnegative, but $\widetilde f_1(x)$ is not.
We have displayed in the bottom panel of  Figure \ref{pos}
the 5\%, 50\%, and 95\% point-wise quantiles for $\widehat f_1$ (left panel) and $\widehat f_2$ (right panel) over 1000 replications.
We have observed that the 90\% confidence bands of $f_1$ and $f_2$ capture the corresponding true densities.

%
%

In addition, we are not able to compare the results of our method with those by Ma et~al. (2011) as the corresponding implemented algorithm is not publicly available yet. However, we observe that in this specific example the method by that paper leads to similar density estimates as the simple method. In particular, with straightforward mathematical manipulations, one can show that the estimate of $f_2(x)$ by that method is exactly the same as that by the simple method, whereas the estimate of $f_1(x)$ is given by
\begin{eqnarray*}
\widetilde f_1'(x) = \frac{\widetilde r(x) - 0.323\widetilde f_2'(x)}{0.677},
\end{eqnarray*}  
with $\widetilde f_2(x)$ being the kernel density based on $X_{n_1+1}, \ldots, X_{n}$, but using the same bandwidth as $\widetilde r(x)$; in contrast, the corresponding term in the simple method is $\widetilde f_2(x)$ whose bandwidth is chosen based on $X_{n_1+1}, \ldots, X_{n}$.

\section{Real Data Example} \label{section-read-data}

We consider the malaria data described by Vounatsou et~al.~(1998). The data come from a cross-sectional survey of parasitemia and fever of children less than a year old in a village in the Kilombero district of Tanzania (Kitua et~al.~1996). They considered a subset of this data for children of between six and nine months collected in two seasons: (1) January--June, the wet season, when malaria prevalence is high; (2) July--December, the dry season, when malaria prevalence is low. The data sets are available from {\tt http://www.blackwellpublishers.\\co.uk/rss}. We use one of these data sets, which has also been analyzed by Qin and Leung (2005) and Yu et~al.~(2014) with other statistical methods.

The measurements are the parasite levels (per $\mu l$), ranging from 0 to $399952.1$.
Among these measurements, there are $n_1=211$ observations
with positive parasite levels from the mixture sample and
$n_2=81$ observations with positive parasite levels for nonmalaria cases
in the community. Therefore, if we denote these parasite levels (after log transformation) as $X_1, \ldots, X_{n_1}, X_{n_1+1}, \dots, X_n$ with $n=n_1+n_2$, then
\begin{eqnarray*}
X_i|\alpha_i \sim  \alpha_if_1(x) + (1-\alpha_i) f_2(x), \nonumber
\end{eqnarray*}
where $f_1(x)$ and $f_2(x)$ are the pdfs of the log parasite levels
for the malaria and nonmalaria subjects respectively; $\alpha_i$ is the probability that the $i$th subject is a malaria patient. Clearly, when $i>n_1$, $\alpha_i = 0$ as it is known that all the subjects in this group are nonmalaria patients. When $i\le n_1$, $\alpha_i\approx 0.677$ estimated from the proportional of the malaria patients over the fevered patients
in the endemicity and the community (Qin and Leung~2005). Therefore,
\begin{eqnarray*}
X_i|\alpha_i \sim f_2(x) \quad &\mbox{when }& i>n_1\\
X_i|\alpha_i \sim 0.677 f_1(x)+0.323f_2(x) \quad &\mbox{when }& i\le n_1.
\end{eqnarray*}

\begin{figure}[!h]
\caption{Component density estimates for malaria data based on the proposed method and
the simple method described in Section \ref{section-sim}.
}
\centerline{ \includegraphics[scale=0.5,angle=-90]{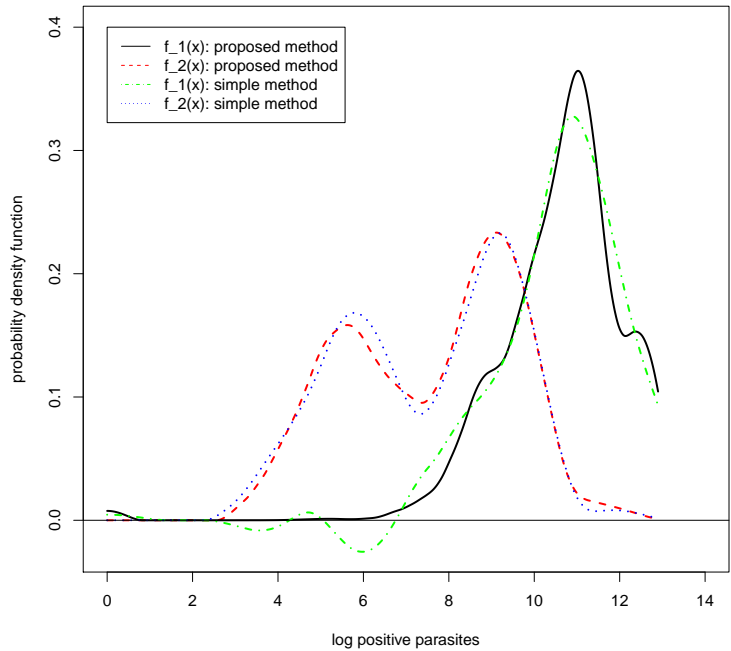}}
\label{figure-real-data1}
\end{figure}

\begin{figure}[!h]
\caption{Histograms for the nonmalaria sample [i.e., that for $f_2(x)$] and the mixture sample [i.e., that for $0.677 f_1(x)+0.323f_2(x)$] along with the corresponding density estimates based on the proposed method.
}
\centerline{ \includegraphics[scale=0.5,angle=-90]{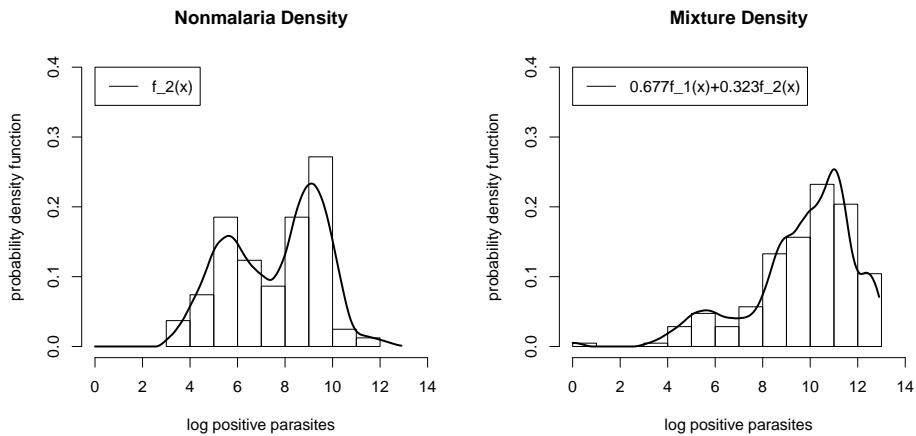}}
\label{figure-real-data2}
\end{figure}

We apply our method and the simple method described in Section \ref{section-sim} on $\{X_i, \balpha_i\}_{i=1}^n$ above, where $\balpha_i = (\alpha_i, 1-\alpha_i)^\tau$. Bandwidths are selected by the algorithm in Section \ref{section-band-sel}, and we get $h_1=0.832$ and $h_2=1.127$. The resultant density estimates $\widehat f_1(x)$, $\widehat f_2(x)$, $\widetilde f_1(x)$, and $\widetilde f_2(x)$
are diplayed in Figure \ref{figure-real-data1}.
Both the ``hat" and ``tilde" estimates for $f_1$ (and $f_2$) are similar in shape.
But $\widetilde f_1(x)$ is not always nonnegative.
Together with the observations in our simulation studies,  we expect that $\widehat f_1(x)$ and $\widehat f_2(x)$ are more efficient than $\widetilde f_1(x)$ and $\widetilde f_2(x)$.
We have also displayed the histograms for the nonmalaria sample [i.e., that for $f_2(x)$] and the mixture sample [i.e., that for $0.677 f_1(x)+0.323f_2(x)$] with the corresponding density estimates from our method in Figure \ref{figure-real-data2}.
From this figure, we observe that our density estimates agree very well with the observed data (see the histogram of the observations from the respective sample).

Furthermore, from Figure \ref{figure-real-data1}, we observe that the density estimate for the log parasite levels of the malaria patients (the black solid line) has a clearer peak and more concentrated curve (centered around 11) than that for the nonmalaria sample (the red dashed line), which has a bimodal feature. From practical point of view, we argue that such an observation is not surprising: the log parasite levels for the nonmalaria sample may be resulted from more than one cause; these causes may lead to different parasite levels and therefore the corresponding density is in fact a mixture of a number of subpopulations. In contrast, the cause for the malaria sample is clear, i.e., the malaria disease; therefore, the resultant density is concentrated and has a clear peak.

\setcounter{equation}{0}
\setcounter{section}{0}
\renewcommand{\theequation} {A.\arabic{equation}}
\section*{Appendix A: Proof of Theorems \ref{theorem-1}--\ref{theorem-1-added-2}}

\subsection*{Proof of Theorem \ref{theorem-1}}
The proof of this theorem uses a similar strategy as that in Levine et.~al. (2011).
Recall that for $(f_1,\ldots,f_M)\in \mathC$,
$w_{i,j}=\frac{\alpha_{i,j}\cn_{h_j}f_j(X_i)}{\sum_{k=1}^M\alpha_{i,k}\cn_{h_k} f_k(X_i)}$.
Then for every $i=1,\ldots,n$, $\sum_{j=1}^M w_{i,j}$ = 1.  By the concavity of the logarithm function, we have for every $(g_1, \ldots,g_M) \in \mathC$,
\begin{eqnarray}
&&l_n(g_1, \ldots, g_M) - l_n(f_1, \ldots, f_M) \nonumber \\
&=& \sum_{i = 1}^n \log\frac{\sum_{j=1}^M \alpha_{ij} \cn_{h_j} g_j(X_i)}{\sum_{j=1}^M \alpha_{ij} \cn_{h_j} f_j(X_i)} \nonumber \\
&=& \sum_{i=1}^n \log \sum_{j=1}^M w_{i,j} \frac{\cn_{h_j}g_j(X_i)}{\cn_{h_j}f_j(X_i)}  \nonumber \\
&\geq & \sum_{i=1}^n \sum_{j=1}^M w_{i,j} \left\{ \log \cn_{h_j}g_j(X_i) - \log \cn_{h_j}f_j(X_i) \right\} \nonumber \\
&=& \sum_{j=1}^M\left\{b_{j}(g_1, \ldots, g_M) - b_{j}(f_1, \ldots, f_M)\right\}, \label{thm-1-1}
\end{eqnarray}
where
\begin{eqnarray}
 b_{j}(g_1, \ldots, g_M) &=& \sum_{i=1}^n  w_{i,j} \log \cn_{h_j}g_j(X_i)  \nonumber \\
 &=& \int \sum_{i=1}^n w_{i,j} K_h(u-X_i) \log g_j(u) du, \label{thm-1-2}
\end{eqnarray}
which is maximized when $g_j(x) = \frac{\sum_{i=1}^n w_{i,j} K_h(x-X_i)}{\sum_{i=1}^n w_{i,j}} = f_j^ \mathG(x)$. This together with (\ref{thm-1-1}) completes the proof of this theorem. \epf

\subsection*{Proof of Theorem \ref{theorem-1-added-1}}

We first show necessity.  Assume $l_n(\widehat f_1,\ldots,\widehat f_M)
=\sup_{(f_1,\ldots,f_M) \in \mathC} l_n(f_1,\ldots,f_M)$. Based on Theorem \ref{theorem-1}, we immediately have $l(\widehat f_1, \ldots, \widehat f_M) = l(\mathG(\widehat f_1, \ldots, \widehat f_M))$. Next we show that $(\widehat f_1, \ldots, \widehat f_M) = \mathG(\widehat f_1, \ldots, \widehat f_M)$.

With exactly the same calculation as (\ref{thm-1-1}) and (\ref{thm-1-2}), we have
\begin{eqnarray*}
0 &=& l(\mathG(\widehat f_1, \ldots, \widehat f_M)) - l(\widehat f_1, \ldots, \widehat f_M) \nonumber \\
&\ge & \sum_{j=1}^M\left\{ \left(\sum_{i=1}^n \widehat w_{i,j} \right)\int \widehat f_j^{\mathG}(x) \log \frac{\widehat f_j^{\mathG}(x)}{\widehat f_j(x)} dx \right\}, \label{thm-1-added-1-1}
\end{eqnarray*}
where $\widehat f_j^{\mathG}$ denotes the $j$th component of $\mathG(\widehat f_1, \ldots, \widehat f_M)$, $\widehat w_{i,j}=\frac{\alpha_{i,j}\cn_{h_j}\widehat f_j(X_i)}{\sum_{k=1}^M\alpha_{i,k}\cn_{h_k} \widehat f_k(X_i)}$. On the other hand, as  $\widehat f_j^{\mathG}$ and $\widehat f_j$ are pdfs, we have
\begin{eqnarray*}
\int \widehat f_j^{\mathG}(x) \log \frac{\widehat f_j^{\mathG}(x)}{\widehat f_j(x)} dx \ge 0.
\end{eqnarray*}
Furthermore for every $j=1,\ldots, M$, since $\sum_{i=1}^n \alpha_{i,j} > 0$ and $(\widehat f_1, \ldots, \widehat f_M)\in \mathC$,  we have $\sum_{i=1}^n \widehat w_{i,j} > 0$. Therefore,
\begin{eqnarray*}
\int \widehat f_j^{\mathG}(x) \log \frac{\widehat f_j^{\mathG}(x)}{\widehat f_j(x)} dx = 0,
\end{eqnarray*}
which together with the continuity of $\widehat f_j(x)$ and $\widehat f_j^{\mathG}(x)$, and the fact that $\log(\cdot)$ is strictly concave leads to $\widehat f_j^{\mathG}(x) = \widehat f_j(x)$ for every $x\in S_{x}$. That is $(\widehat f_1, \ldots, \widehat f_M) = \mathG(\widehat f_1, \ldots, \widehat f_M)$ as claimed before.

We proceed to show the sufficiency. Assume $(\widehat f_1, \ldots, \widehat f_M) = \mathG(\widehat f_1, \ldots, \widehat f_M)$.
Let $\widehat\bbf=(\widehat f_1, \ldots, \widehat f_M)$.
 For an arbitrary $\bbf = (f_1,\ldots,f_M) \in \mathF_n$,
 we need to show that $l_n(\bbf )\leq l_n(\widehat \bbf)$.

Define
\begin{eqnarray}
H(t) = l_n(\widehat \bbf + t(\bbf - \widehat \bbf)), \label{def-H(t)}
\end{eqnarray}
with $t\in [0, 1]$. Next, we verify that $H(\cdot)$ has the following properties.
\begin{itemize}
\item[(P1).] $H(t)$ is a concave function in $[0,1]$.
\item[(P2).] $H(t)$ is continuously differentiable in $(0,1)$, $H'(0+)$ exists, and $H'(0+) = 0$.
\end{itemize}
We first show (P1) above. Note that $l_n$ is concave in $\mathC$, we immediately have for every $t_1, t_2 \in [0, 1]$,
\begin{eqnarray*}
H\left(\frac{t_1+t_2}{2}\right) &=& l_n\left(\frac{\left\{\widehat \bbf + t_1(\bbf - \widehat \bbf)\right\} + \left\{\widehat \bbf + t_2(\bbf - \widehat \bbf)\right\}}{2} \right) \nonumber \\
&\ge& \frac{1}{2}l_n\left(\widehat \bbf + t_1(\bbf - \widehat \bbf) \right) + \frac{1}{2}l_n\left(\widehat \bbf + t_2(\bbf - \widehat \bbf) \right) \nonumber \\
&=& \frac{1}{2} H(t_1) + \frac{1}{2} H(t_2),
\end{eqnarray*}
leading to (P1). We proceed to show (P2). First, to verify that $H(t)$ is continuously differentiable in $(0,1)$ and the existence of $H'(0+)$, it suffices to verify that for every $x\in S_x$ and $j=1,\ldots,M$, $\int K_h(u-x)\log \left[\widehat f_j(u) + t\left\{f_j(u) - \widehat f_j(u)\right\}\right] du$ is continuously differentiable when $t\in (0, 1)$, right differentiable at $t=0$, and the derivative can be exchanged with the integration. This is valid because of the definition of $\mathF_n$ and the dominant convergence theorem. Therefore, it is left to verify $H'(0+) = 0$.
For notational convenience, we denote $\bbf_t = \widehat \bbf + t(\bbf - \widehat \bbf) =  (f_{1, t}, \ldots, f_{M, t})$ and let $(f_{1,t}^{\mathG}, \ldots, f_{M,t}^{\mathG}) = \mathG(f_{1,t}, \ldots, f_{M,t})$. Using the chain rule of derivatives, we have for every $t\in (0,1)$,
\begin{eqnarray*}
H'(t) &=& \sum_{i=1}^n\sum_{j=1}^M\frac{\alpha_{i,j}\cn_{h_j} f_{j,t} }{\sum_{k=1}^M\alpha_{i,k}\cn_{h_k} f_{k,t}(X_i)}\int \frac{K_{h_j}(u-X_i)}{f_{j,t}(u)}\left\{f_j(u) - \widehat f_j(u)\right\} du \nonumber \\
&=& \sum_{j=1}^M \int \frac{f_{j,t}^{\mathG}(u)}{f_{j,t}(u)}\left\{f_j(u) - \widehat f_j(u)\right\} du.
\end{eqnarray*}
Noting the fact that $f_{j,0} =  \widehat f_j$ and $\widehat f_j^{\mathG} = \widehat f_j$ based on our assumption, we immediately have
\begin{eqnarray*}
H'(0+) &=& \sum_{j=1}^M \int \frac{f_{j,0}^{\mathG}(u)}{f_{j,0}(u)}\left\{f_j(u) - \widehat f_j(u)\right\} du\\
&=& \sum_{j=1}^M \int \left\{f_j(u) - \widehat f_j(u)\right\} du = 0,
\end{eqnarray*}
which completes our proof of (P2) above. Now based on (P1) and (P2) and the property of the concave functions, we immediately have
\begin{eqnarray}
H(1)\le H(0) + H'(0+)(1-0), \nonumber
\end{eqnarray}
which is
\begin{eqnarray*}
l_n (\bbf) \le l_n(\widehat \bbf).
\end{eqnarray*}
This completes the proof of the theorem. \epf

\subsection*{Proof of Theorem \ref{theorem-1-added-2}}

Since $(f_1^{s}, \ldots, f_M^{s})\in \mathF_n$, for every $j=1,\ldots, M$, we can write
\begin{eqnarray*}
f_j^s(x) = \frac{\sum_{i=1}^nw_{i,j}^s K_{h_j}(x-X_i)}{\sum_{i=1}^n w_{i,j}^s}.
\end{eqnarray*}
Clearly, for every $s$, the collection of the coefficients $w^s = \{w_{i,j}^s: i=1,\ldots,n; j = 1,\ldots,M\}$ belongs to
\begin{eqnarray*}
\Omega_w = \left\{\{w_{i,j}: i=1,\ldots,n; j = 1,\ldots,M\}: 0 \le w_{i,j} \le 1 \right\},
\end{eqnarray*}
which is a closed subset of $\RR^{nM}$. Therefore, there exists a subsequence of $w^s$, namely $w^{s_l}$, and $w^\infty = \{w_{i,j}^\infty: i=1,\ldots,n; j = 1,\ldots,M\} \in \Omega_w$, such that
\begin{eqnarray}
\lim_{l\to\infty}w^{s_l} = w^\infty. \label{limit.w}
\end{eqnarray}
Let
\begin{eqnarray*}
f_j^\infty(x) = \frac{\sum_{i=1}^nw_{i,j}^\infty K_{h_j}(x-X_i)}{\sum_{i=1}^n w_{i,j}^\infty}.
\end{eqnarray*}
It can be readily checked that
\begin{equation}
\lim_{l\to\infty} f_j^{s_l}(x)=f_j^\infty(x)
\label{limit.f}
\end{equation}
for all $x\in S_{x}$
and hence
\begin{eqnarray*}
\lim_{l\to \infty} l_n (f_1^{s_l}, \ldots, f_M^{s_l})  = l_n (f_1^\infty, \ldots, f_M^\infty),
\end{eqnarray*}
which together with Theorem \ref{theorem-1} ensures
\begin{eqnarray*}
\lim_{s\to \infty} l_n (f_1^{s}, \ldots, f_M^{s})  = l_n (f_1^\infty, \ldots, f_M^\infty).
\end{eqnarray*}
It is left to show
\begin{eqnarray}
\label{thm-1-added-needed}
\mathG(f_1^\infty, \ldots, f_M^\infty) = (f_1^\infty, \ldots, f_M^\infty).
\end{eqnarray}
Then based on Theorem \ref{theorem-1-added-1}, we have
$$
 l_n (f_1^\infty, \ldots, f_M^\infty)
 =l_n(\widehat f_1,\ldots,\widehat f_M),
$$
which completes our proof of this theorem.

In fact, along the subsequence $s_l$ defined above, using the same derivations as (\ref{thm-1-1}) and (\ref{thm-1-2}), we have
\begin{eqnarray*}
\nonumber
0&=& \lim_{l\to \infty}\left\{l_n(f_1^{s_l+1}, \ldots, f_M^{s_l+1})- l_n(f_1^{s_l}, \ldots, f_M^{s_l})\right\}\nonumber \\
&\ge&  \lim_{l\to \infty} \sum_{j=1}^M\left\{ \left(\sum_{i=1}^n w_{i,j}^{s_l} \right)\int f_j^{s_l+1}(x) \log \frac{f_j^{s_l+1}(x)}{f_j^{s_l}(x)} dx \right\}\geq0.
\end{eqnarray*}
Hence
\begin{equation}
\lim_{l\to \infty} \sum_{j=1}^M\left\{ \left(\sum_{i=1}^n w_{i,j}^{s_l} \right)\int f_j^{s_l+1}(x) \log \frac{f_j^{s_l+1}(x)}{f_j^{s_l}(x)} dx \right\}=0.
\label{limit.diff}
\end{equation}
On the other hand, note that (\ref{limit.f}) implies
$\lim_{l\to\infty} \mathG(f_1^{s_l}, \ldots, f_M^{s_l}) = \mathG(f_1^\infty, \ldots, f_M^\infty)$, or equivalently,
\begin{equation}
 \lim_{l\to \infty} (f_1^{s_{l}+1}, \ldots, f_M^{s_{l}+1}) = (f_1^{\infty,\mathG}, \ldots, f_M^{\infty,\mathG}),
 \label{limit.mathg}
 \end{equation}
 where $(f_1^{\infty,\mathG}, \ldots, f_M^{\infty,\mathG})=\mathG(f_1^\infty, \ldots, f_M^\infty)$.
 Combining (\ref{limit.w}), (\ref{limit.f}), (\ref{limit.diff}), and (\ref{limit.mathg}), we have
$$
\sum_{j=1}^M\left\{ \left(\sum_{i=1}^n w_{i,j}^{\infty} \right)\int f_j^{\infty,\mathG}(x) \log \frac{f_j^{\infty,\mathG}(x)}{f_j^{\infty}(x)} dx \right\} = 0,
$$
which indicates for every $j=1,\ldots, M$,
\begin{eqnarray}
\int f_j^{\infty,\mathG}(x) \log \frac{f_j^{\infty,\mathG}(x)}{f_j^{\infty}(x)} dx  = 0.
\label{limit.last}
\end{eqnarray}
With the continuity of $f_j^{\infty}(x)$ and $f_j^{\infty,\mathG}(x)$, and the fact that $\log(\cdot)$ is strictly concave, (\ref{limit.last}) implies $f_j^{\infty}(x)=f_j^{\infty,\mathG}(x)$. That is
\begin{eqnarray*}
\mathG(f_1^\infty, \ldots, f_M^\infty) = (f_1^\infty, \ldots, f_M^\infty),
\end{eqnarray*}
which proves (\ref{thm-1-added-needed}), and therefore completes the proof of this theorem. \epf

\setcounter{equation}{0}
\setcounter{section}{0}
\renewcommand{\theequation} {B.\arabic{equation}}
\section*{Appendix B: Proof of Theorems \ref{consistency-theorem-1} -- \ref{theorem-4}}

\subsection*{Technical Conditions}

We impose the following conditions to facilitate our technical developments for Theorems \ref{theorem-3} and \ref{theorem-4}. They are not necessarily the weakest possible.

\begin{itemize}
\item[] \underline{Condition 1}:  There exists a bandwidth $h$ such that   $C_1 \le \inf_{n,j} h_j/h \le \sup_{n, j}h_j/h \le C_2$, where $C_1> 0$ and $C_2 >0$ are universal constants. Furthermore, $h\to 0$ and $nh \to \infty$ when $n\to \infty$.

\item[]   \underline{Condition 2:}  The kernel function $K(x)$ is symmetric about 0, supported and continuous on $[-L, L]$ for some $L>0$  and $\inf_{x\in[-L, L]} K(x) >0$.
The $a$th-order derivative $K^{(a)}(x)$ of $K(x)$ exists for every $a = 1,2, \ldots$ and $x\in(-L, L)$.
Further $\sup_{a,x}|K^{(a)}(x)|$ is bounded.

\item[] \underline{Condition 3:}
The true component pdfs $f_{0,j}(x)$, $j=1,\ldots,M$ are supported on $S_{x} = [c_1, c_2]$ and are twice continuously differentiable in $(c_1,c_2)$ with bounded second order derivatives. Furthermore,  $\inf_{x\in S_x}f_{0,j}(x) > 0$.

\item[] \underline{Condition 4}:
Let $S_{\gamma}$ be the support for $\gamma(\balpha)$. There exists $M\times 1$ vectors $\balpha_{0,1}, \ldots, \balpha_{0,M}$ in $S_\gamma$ satisfy (i) and (ii) below.
\begin{itemize}
\item[(i).] The $M$ vectors $\balpha_{0,1}, \ldots, \balpha_{0,M}$ are linearly independent.
\item[(ii).] There exist balls $\mathO_j\subset S_\gamma, j=1,\ldots, M$, $\balpha_{0,j} \in \mathO_j$, $\mathO_j$ are disjoint, and $\gamma(\balpha) > 0$ for every $\balpha \in \mathO_j$.
\end{itemize}

Condition 1 requires that the $M$ bandwidths have the same order. Condition 2 requires that the kernel function $K(x)$ is symmetric and is sufficiently smooth.
Condition 3 requires the component pdfs are sufficiently smooth and
is positive on the support of $X$.
Condition 4 is a identifiability condition, which is satisfied when $\balpha$ is a continuous random vector,  or a discrete random vector with at least $M$ supports.

\end{itemize}

\subsection*{Preliminary preparation}
 The proof of Theorems \ref{consistency-theorem-1}--\ref{theorem-4}
 heavily relies on the well developed results for the M-estimation in empirical process.
 We use van der Vaart and Wellner (1996) (VM) as the main reference and adapt the commonly used notation in this book.
 In this section, we introduce some necessary notation and review two important results.

We first review some notation  necessary for  introducing the result for the M-estimation.
Let $``\lesssim"$ ($``\gtrsim"$) denote smaller (greater) than, up to a universal constant. Throughout, we will use $C$ to denote a sufficiently large universal constant.
For a set $\mathM$ of functions of $(x,\balpha)$, we define
\begin{eqnarray}
\GG_n m&=&\sqrt{n}\left\{\frac{1}{n} \sum_{i=1}^n  m(X_i,\balpha_i) - E_0 m(X,\balpha) \right\} \mbox{ for }m\in \mathM\label{def-GG-n1}\\
||\GG_n||_{\mathM}&=&\sup_{m\in \mathM}\left| \GG_n m \right|.
\label{def-GG-n2}
\end{eqnarray}
Here $E_0$ means the expectation is taken under $\gamma(\balpha)\widetilde p_0(x,\balpha)$. This convention will be used throughout the proof.
The Hellinger distance between two non-negative functions $m_1(x,\balpha)$ and $m_2(x,\balpha)$ is defined to be
\begin{eqnarray*}
d(m_1, m_2) = \left[\int_{S_{\gamma}}\int_{\RR} \left\{m_1^{1/2}(x,\balpha) - m_2^{1/2}(x,\balpha) \right\}^2 dxd\balpha \right]^{1/2}.
\end{eqnarray*}

Let $\mathP_n$ denote the class of functions:
\begin{eqnarray}
\mathP_n &=& \left\{p(x, \balpha) = \sum_{j=1}^M\alpha_{j}\cn_{h_j} f_j(x): (f_1, \ldots, f_M) \in \mathF_n \right\},\label{def-mathP-n}
\end{eqnarray}
where $\mathF_n$ is defined by (\ref{def-mathF-n}).
For any nonnegative functions $p(x,\balpha)$ and $p_1(x,\balpha)$, we define
\begin{eqnarray}
m_{p, p_1}(X, \balpha) &=& \log\frac{p(X, \balpha)+p_1(X, \balpha)}{2p_1(X, \balpha)}; \label{def-m-p-p}\\
\MM_n(p, p_1) &=&\frac{1}{n}\sum_{i=1}^n m_{p,p_1}(X_i, \balpha_i); \label{def-MM-n}\\
M_n(p, p_1) & = &E_0 m_{p, p_1}(X, \balpha);\label{def-M-n}\\
\mathM_{n, \delta, p, p_1} &=& \left\{m_{p,p_1}-m_{p_1, p_1}: p\in \mathP_n, d(\gamma p, \gamma p_1) < \delta\right\}. \label{def-mathM-n}
\end{eqnarray}

With the above preparation, we present an important lemma,
which is an application of Theorem 3.4.1 of van der Vaart and Wellner (1996) to our current setup.
It serves the basis for our subsequent proof.
\begin{lemma}
\label{lemma0.vm}
Suppose the notation $\MM_n$, $M_n$,  and $\|\GG_n\|_{\mathM_{n, \delta, p, \widetilde p_0}}$ are defined above,
 $\widetilde p_0(x, \balpha) = \sum_{j=1}^M \alpha_{j} f_{0,j}(x)$ is the true conditional density of $X$ given $\balpha$, and $\gamma(\cdot)$ is the marginal density of $\balpha$.
If the following three conditions are satisfied:
\begin{enumerate}
\item[(C1)] for every $n$ and $p\in \mathF_n$,
$
M_n(p,\widetilde p_0)-M_n(\widetilde p_0,\widetilde p_0)\lesssim -d^2(\gamma p,\gamma\widetilde p_0);
$
\item[(C2)] for every $n$ and $\delta>0$, $E_0 \|\GG_n\|_{\mathM_{n, \delta, p, \widetilde p_0}} \lesssim \phi_n(\delta) $ for functions $\phi_n$ such that $\phi_n(\delta)/\delta^\alpha$ is decreasing on $(0,\infty)$ for some $\alpha<2$;
\item[(C3)] $\MM_n(\widehat p,\widetilde p_0)\geq \MM_n(\widetilde p_0,\widetilde p_0)-O_p(r_n^{-2})$, where
$\widehat p(x,\balpha)= \sum_{j=1}^M \alpha_j \cn_{h_j} \widehat f_{j}(x)$ and $r_n $ satisfies
$
r_n^2\phi(1/r_n)\leq \sqrt{n},
$
for every $n$;
\end{enumerate}
then
$$
r_n d (\gamma\widehat p,\gamma\widetilde p_0)=O_p(1).
$$
\end{lemma}

An difficult step in the application of the above lemma is to verify Condition C2.
An useful technique is to establish a connection between
$E_0 \|\GG_n\|_{\mathM_{n, \delta, p, \widetilde p_0}}$ and the bracketing integral of
the class $\gamma \mathP_n$.
For the convenience of presentation in next subsections, we introduce some necessary notation and
review an important lemma.

We first introduce the concept of bracketing numbers, which will be used to define the bracketing integral.
Consider a set $\mathM$ of functions and the norm $\|\cdot\|$ defined on the set $\mathM$.
For any $\epsilon>0$, the bracketing number $N_{[]}(\epsilon,\mathM,\|\cdot\|)$ is
the minimum number of $N$ for which there exists a set of pairs of functions $\{(l_j,u_j)\}_{j=1}^N$
such that (i) $\| u_j-l_j\|<\epsilon$ and (ii) for any $m\in \mathM$, there exists a $j=j(m)$ such that
$l_j\leq m\leq u_j$.
The bracketing integral of the class $\mathM$ is then defined to be
\begin{equation}
\label{bracket.integral}
\widetilde J_{[]}(\delta, \mathM, \|\cdot\|) = \int_0^\delta\sqrt{1+\log N_{[]}(\epsilon, \mathM, \|\cdot\|)}d\epsilon.
\end{equation}

Next, we review a result about the covering number of a class of continuous functions, which
will be useful to calculate the bracketing number of $\gamma \mathP_n$
and the bracketing integral of $\gamma \mathP_n$.
For every function $f$ defined on $\mathA\subset \RR$ and a positive integer $a$, define the norm
\begin{eqnarray*}
\|f\|_{a} = \max_{k: k \le a} \sup_{x}|f^{(k)}(x)|
\end{eqnarray*}
where the suprema are taken over $x\neq y$ in the interior of $\mathA$; $f^{(k)}(x)$ denotes the $k$th order derivative of $f$; $f^{(0)} = f$. Let $C_W^{a}(\mathA)$ be the set of all continuous functions $f:\mathA \mapsto \RR$ with $\|f\|_a \le W$.

\begin{lemma} \label{lemma-1}
Let $\mathA$ be a length $L$ interval in $\RR$. There exists a constant $K<\infty$ depending only on $a$ and $L$ such that
\begin{eqnarray*}
\log N_{[]}(\epsilon, C_1^a(A), L_r(Q)) \le K/\epsilon^{1/a},
\end{eqnarray*}
for every $r\ge 1$, $\epsilon >0$, and any probability measure $Q$ on $\RR$.
Here $L_r(Q)$ is the $L_r$-norm under the probability measure $Q$.
\end{lemma}
This lemma is the special case of the Corollary 2.7.2 of VW; see Page 157.

\subsection*{Proof of Theorem \ref{consistency-theorem-1}: Consistency of $d(\gamma \widehat p, \gamma \widetilde p_0)$}

In this section, we show Theorem \ref{consistency-theorem-1}, which establishes the consistency of $d(\gamma \widehat p, \gamma \widetilde p_0)$ and plays a key role in the proofs of Theorems \ref{theorem-3} and \ref{theorem-4} subsequently. Recall that we need to show
\begin{eqnarray*}
d(\gamma \widehat p, \gamma\widetilde p_0) = O_p(h^{0.5}) + O_p(n^{-0.5+\vartheta}h^{-0.5}).
\end{eqnarray*}

This proof  contains three steps. In each step, we verify one condition in Lemma \ref{lemma0.vm}.

In {\it Step 1}, we verify that Condition C1 in Lemma \ref{lemma0.vm} is satisfied. We need the following lemma regarding the property of smoothing operator $\cn_h$.
\begin{lemma} \label{proposition-1}
Consider $\cn_h f(x)$ defined by (\ref{def-cn}), then for any density function $f(x)$, we have
\begin{eqnarray*}
\int_{\RR} \cn_hf(x) dx \le 1.
\end{eqnarray*}
\end{lemma}

\pf By the concavity of the logarithm and Jensen's inequality, we have
\begin{eqnarray*}
\int_{\RR} \cn_h f(x) dx &=& \int_{\RR} \exp\left\{\int_{\RR} K_h(u-x)\log f(u) du\right\}dx  \nonumber \\
&\leq& \int_{\RR} \int_{\RR} K_h(u-x)f(u) du dx\\
&=&  \int_{\RR} f(u)\int_{\RR} K_h(u-x) dx du=1. \qed
\end{eqnarray*}
We now move back to verify Condition C1.
For any $p\in \mathF_n$, let $q=(p+\widetilde p_0)/2$.
Since $\log x\le 2(\sqrt{x} - 1)$ for every $x>0$,  we have that
\begin{eqnarray*}
M_n(p,\widetilde p_0)-M_n(\widetilde p_0,\widetilde p_0)&=&E_0\left(\log\frac{q}{\widetilde p_0}\right)\\
 &\le& 2E_0\left( \frac{q^{1/2}}{\widetilde p_0^{1/2}} -1\right) \nonumber \\
&=& -d^2(\gamma \widetilde p_0, \gamma q) + \int_{\RR} \gamma(q-\widetilde p_0) dx d\balpha \nonumber \\
&=& -d^2(\gamma \widetilde p_0, \gamma q) + 0.5\int_{\RR} \gamma \left\{\int_{\RR} pdx -1 \right\} d\balpha \nonumber\\
&\le & -d^2(\gamma \widetilde p_0, \gamma q),
\end{eqnarray*}
where, to achieve the last ``$\leq$", we have applied Lemma \ref{proposition-1}.
Note that
$$
\left|
\sqrt{\gamma p}-\sqrt{\gamma \widetilde p_0}
\right|
=2\frac{\sqrt{\gamma q}+\sqrt{\gamma \widetilde p_0}}{\sqrt{\gamma p}+\sqrt{\gamma \widetilde p_0}}
\left|
\sqrt{\gamma q}-\sqrt{\gamma \widetilde p_0}
\right|\leq 4 \left|
\sqrt{\gamma q}-\sqrt{\gamma \widetilde p_0}
\right|,$$
which implies that
$$
-d^2(\gamma \widetilde p_0, \gamma q)\leq -\frac{1}{16}d^2(\gamma \widetilde p_0, \gamma p).
$$
Therefore
$$
M_n(p,\widetilde p_0)-M_n(\widetilde p_0,\widetilde p_0)\leq -\frac{1}{16}d^2(\gamma \widetilde p_0, \gamma p).
$$
Hence Condition C1 of Lemma \ref{lemma0.vm} is satisfied.


%
%
%

In {\it Step 2},  we establish the upper bound for  $E_0 \|\GG_n\|_{\mathM_{n, \delta, p, \widetilde p_0}}$.
Following exactly the same lines as that of  Theorem 3.4.4 in VM, we get that
\begin{eqnarray}
E_{0} \|\GG_n\|_{\mathM_{n, \delta, p, \widetilde p_0}} \lesssim \widetilde J_{[]}(\delta, \gamma\mathP_{n}, d) \left\{1+\frac{\widetilde J_{[]}(\delta, \gamma\mathP_{n}, d)}{\delta^2 \sqrt{n}}\right\}, \label{max-cond-2}
\end{eqnarray}
where  the bracketing integral $\widetilde J_{[]}$ is defined in (\ref{bracket.integral}).
Lemma \ref{entropy} below gives the upper bound for $\widetilde J_{[]}(\delta, \gamma\mathP_{n}, d)$, which, combined with (\ref{max-cond-2}), immediately leads to $\phi_n(\cdot)$ in Condition C2 of Lemma \ref{lemma0.vm}.

\begin{lemma} \label{entropy}
Let $a$ be an arbitrary positive integer. Then
\begin{eqnarray}
\widetilde J_{[]}(\delta, \gamma\mathP_{n}, d) \lesssim \delta^{1-1/(2a)}\sum_{j=1}^M (\log h_j)^{0.5} h_j^{-0.5-0.25/a}.  \label{max-cond-3}
\end{eqnarray}
\end{lemma}

\pf
Consider
$$
\mathP_{n,j} = \{\cn_{h_j} f: f = \frac{\sum_{i=1}^nw_{i,j} K_{h_j}(x-X_i)}{\sum_{i=1}^n w_{i,j}}; 0 \le w_{i,j} \le 1\}.
$$
Let $S_x^*=[c_1-\Delta,c_2+\Delta]$, where $\Delta>0$ is an arbitrarily small constant.
Note that for any $ g\in\mathP_{n,j}$, $g(x)=0$ when $x\notin S_x^*$.
In the following proof, we focus on the function class defined on $ S_x^*$.

With Condition 2, we first check that for any arbitrary $a >0$, we have
\begin{eqnarray}
\left(\frac{h_j}{\log h_j}\right)^a \sqrt{h_j} C_2\sqrt{\mathP_{n,j}} \subset C_1^a(S_x^*)
 \end{eqnarray}
 for some universal constant $C_2 > 0$.
 Here $S_x^*=[c_1-\Delta,c_2+\Delta]$, where $\Delta>0$ is an arbitrarily small constant.
 For presentational brevity, we only show the case of $a=1$; the cases of $a=2,3,\ldots,$ can be proved similarly. For any $\sqrt{N_{h_j}f} \in \sqrt{\mathP_{n,j}}$, by straightforward calculus, we have
\begin{eqnarray}
\left(\sqrt{N_{h_j}f}\right)' = \frac{1}{h_j} \exp\left\{0.5\int_{\RR} K(t) \log f(x+th_j) dt\right\} \int_{\RR} K'(t)\log f(x+th_j)dt. \label{lem-en-2}
\end{eqnarray}
For any function $f(x)$, let $f^+(x)=\max\{f(x),0\}$ and $f^-(x)=\max\{-f(x),0\}$ denote the positive and negative parts of $f(x)$, respectively.
Using the conditions that $K(t)$ is bounded below and $|K'(t)|$ is bounded in Condition C2, we further have
\begin{eqnarray*}
\left|\left(\sqrt{N_{h_j}f}\right)'\right| &\lesssim& \frac{1}{h_j} \exp\left\{0.5\int_{\RR} K(t) \log f(x+th_j) dt\right\} \int_{\RR} K(t)\left| \log f(x+th_j)\right|dt\\
&\leq&  \frac{1}{h_j} \exp\left[0.5\int_{\RR} K(t)\left\{ \log f(x+th_j)\right\}^+ dt -0.5\int_{\RR} K(t)\left\{ \log f(x+th_j)\right\}^- dt\right] \\
&&\times \left[\int_{\RR} K(t)\left\{ \log f(x+th_j)\right\}^+dt+\int_{\RR} K(t)\left\{ \log f(x+th_j)\right\}^-\right]dt.
\end{eqnarray*}
Note that $\int_{\RR} K(t)\left\{ \log f(x+th_j)\right\}^+dt\lesssim \log(1/h_j)$. Hence
\begin{eqnarray*}
\left|\left(\sqrt{N_{h_j}f}\right)'\right| &\lesssim& \frac{1}{h_j^{1.5}} \exp\left[-0.5\int_{\RR} K(t)\left\{ \log f(x+th_j)\right\}^- dt\right]\\
&&\times \left[\log(1/h_j)+\int_{\RR} K(t)\left\{ \log f(x+th_j)\right\}^-\right]dt.
\end{eqnarray*}
Note that for any $x\geq0$, $x\exp(-0.5x)<1$.
Then
\begin{eqnarray*}
\left|\left(\sqrt{N_{h_j}f}\right)'\right| &\lesssim& \frac{1}{h_j^{1.5}} \log(1/h_j)+\frac{1}{h_j^{1.5}}
\lesssim\frac{1}{h_j^{1.5}} \log(1/h_j).
\end{eqnarray*}

Now, by Lemma \ref{lemma-1} and view $d$ on $\frac{h_j^{2a+1}}{(\log h_j)^{2a}}C_2^2\mathP_{n,j}$ as the $L_2$-distance on $\frac{h_j^{a+0.5}}{(\log h_j)^{a}}C_2\sqrt{\mathP_{n,j}}$, we have
\begin{eqnarray*}
\log N_{[]}\left(\epsilon, \frac{h_j^{2a+1}}{(\log h_j)^{2a}}C_2^2\mathP_{n,j}, d\right) = \log N_{[]}\left(\epsilon, \frac{h_j^{a+0.5}}{(\log h_j)^{a}}C_2\sqrt{\mathP_{n,j}}, L_2\right) \lesssim 1/\epsilon^{1/a}. \label{max-cond-3-1}
\end{eqnarray*}
On the other hand, under $d$, for every $\epsilon$-length bracket of  $\frac{h_j^{2a+1}}{(\log h_j)^{2a}}C_2^2\mathP_{n,j}$, it is a length $\epsilon (\log h_j)^{a}/(h_j^{a+0.5}C_2)$ bracket in $\mathP_{n,j}$. Therefore,
\begin{eqnarray*}
\log N_{[]}\left(\epsilon(\log h_j)^{a}/(h_j^{a+0.5}C_2), \mathP_{n,j}, d\right) = \log N_{[]}\left(\epsilon, \frac{h_j^{2a+1}}{(\log h_j)^{2a}}C_2^2\mathP_{n,j}, d\right) \lesssim 1/\epsilon^{1/a},
\end{eqnarray*}
which immediately implies
\begin{eqnarray}
\log N_{[]}\left(\epsilon, \mathP_{n,j}, d\right) \lesssim \log h_j/\{\epsilon(h_j^{a+0.5})\}^{1/a}. \label{max-cond-3-2}
\end{eqnarray}
For notational simplicity, we write $N_j = N_{[]}(\epsilon, \mathP_{n,j}, d)$. Then for every $j$, there exist a set of $\epsilon$-brackets $\mathB_j = \{[u_{i,j}, v_{i,j}]: i=1,\ldots, N_j\}$ that covers $\mathP_{n,j}$. Let
\begin{eqnarray*}
\mathB = \left\{[p_L(x, \balpha), p_U(x, \balpha)]: p_L = \sum_{j=1}^M \alpha_j u_{i_j, j}(x), p_U = \sum_{j=1}^M \alpha_j v_{i_j,j}(x), \mbox{ for every }j, i_j \in \{1, \ldots, N_j\}\right\}.
\end{eqnarray*}
Clearly, $\mathB$ covers $\gamma\mathP_{n}$ with $\Pi_{j=1}^M N_j$ brackets.

Next we consider the minimum bracket length.
Note that for any $x,x',y,y' \geq 0$, we have
\begin{eqnarray*}
\{(x+y)^{1/2} - (x'+y')^{1/2}\}^2 \le (x^{1/2}-x'^{1/2})^2 + (y^{1/2} - y'^{1/2})^2.
\end{eqnarray*}
Hence for any $[p_L(x, \balpha), p_U(x, \balpha)] \in \mathB$,
\begin{eqnarray*}
d^2(p_L(x, \balpha), p_U(x, \balpha)) \le \sum_{j=1}^Md^2(\alpha_ju_{i_j,j}, \alpha_jv_{i_j,j})
\le \sum_{j=1}^Md^2(u_{i_j,j}, v_{i_j,j}) \le M \epsilon^2.
\end{eqnarray*}
This indicates for every $\epsilon >0$,
\begin{eqnarray*}
\log N_{[]}(\epsilon, \gamma\mathP_{n}, d)  \lesssim \log N_{[]}(\sqrt{M}\epsilon, \gamma\mathP_{n}, d) \le \sum_{j=1}^M\log N_j  \lesssim \sum_{j=1}^M \frac{\log h_j}{\epsilon^{1/a}h_j^{1+0.5/a}}.
\end{eqnarray*}
This proves (\ref{max-cond-3}). \epf

With the help of Lemma \ref{entropy}, we set
$$
\phi_n(\delta) = \delta^{1-1/(2a)}\sum_{j=1}^M (\log h_j)^{0.5} h_j^{-0.5-0.25/a}\left(1+\frac{1}{\sqrt{n} \delta^{1+1/(2a)}}\sum_{j=1}^M (\log h_j)^{0.5}h_j^{-0.5-0.25/a}\right).
$$
Obviously, $\phi_n(\delta)/\delta^\alpha$ with $\alpha=1$ is a decreasing function of $\delta$.
This verifies Condition C2 of Lemma \ref{lemma0.vm}.

In {\it Step 3}, we check
\begin{eqnarray}
\MM_n(\widehat p, \widetilde p_0) \ge \MM_n(\widetilde p_0, \widetilde p_0) + O_p(h). \label{max-cond-4}
\end{eqnarray}

Let $p_n(x, \balpha) = \sum_{j=1}^M\alpha_{j}\cn_{h_j} \mathS_{h_j}f_{0,j}(x)$, where for $j=1,\ldots, M$,
\begin{equation}
\mathS_{h_j} f_{0,j}(x) = \left\{
\begin{array}{cl}
c_{h_j,j} f_{0,j}(c_2),&x\in[c_2, c_2+Lh_j]\\
c_{h_j,j} f_{0,j}(x),&x\in[c_1, c_2]\\
c_{h_j,j} f_{0,j}(c_1),&x\in[c_1-Lh_j, c_1]\\
0,&\mbox{otherwise}
\end{array}
\right.,
\end{equation}
where $c_{h_j,j}$ is a constant such that $\int_{\RR} \mathS_{h_j} f_{0,j}(x) dx=1$.

Note that $\MM_n (\widetilde p_0, \widetilde p_0 )=0$ and $\log(x)$ is concave. We have
\begin{eqnarray*}
\MM_n (\widehat p, \widetilde p_0 )-\MM_n (\widetilde  p_0, \widetilde p_0 ) &=&
\frac{1}{n}\sum_{i=1}^n \log\frac{\widehat p(X_i,\balpha_i) + \widetilde p_0(X_i,\balpha_i)}{2\widetilde p_0(X_i,\balpha_i)}\\
& \ge& \frac{1}{2n}\sum_{i=1}^n\left\{\log   \widehat p(X_i,\balpha_i)- \log \widetilde p_0(X_i,\balpha_i)\right\}   \nonumber \\
&=& \frac{1}{2n}\sum_{i=1}^n\left\{\log   \widehat p(X_i,\balpha_i)- \log p_n(X_i,\balpha_i)\right\}\nonumber\\
&& +\frac{1}{2n}\sum_{i=1}^n\left\{\log p_n(X_i,\balpha_i)- \log \widetilde p_0(X_i,\balpha_i)\right\}
\nonumber\\
&\geq& \frac{1}{2n}\sum_{i=1}^n\left\{\log p_n(X_i,\balpha_i)- \log \widetilde p_0(X_i,\balpha_i)\right\},
\end{eqnarray*}
where the step follows from the fact that
$$
\sum_{i=1}^n\left\{\log   \widehat p(X_i,\balpha_i)- \log  p_n(X_i,\balpha_i)\right\}
=l_n(\widehat f_1,\ldots,\widehat f_M)-l_n(\mathS_{h_1}f_{0,1},\ldots,\mathS_{h_M}f_{0,M})\geq0.
$$

Let
$$
I=\frac{1}{n}\sum_{i=1}^n\left\{\log p_n(X_i,\balpha_i)- \log \widetilde p_0(X_i,\balpha_i)\right\}.
$$
Therefore, to show (\ref{max-cond-4}), we only need to verify that
\begin{eqnarray*}
I = O_p(h),
\end{eqnarray*}
which is valid based on Lemma \ref{lemma-var} below.
\begin{lemma} \label{lemma-var}
Assume Conditions 1--3. We have
\begin{eqnarray}
E_0\log (p_n/\widetilde p_0) &=& O(h) \label{bias-1} \\
\var\left(\log\frac{p_n(X, \balpha)}{\widetilde p_0(X, \balpha)}\right) &=& O(h^2). \label{var-1}
\end{eqnarray}
\end{lemma}
\pf
In the proof, we need the approximation of $\log (p_n/\widetilde p_0) $.
Note that
$$
\log (p_n/\widetilde p_0) =  \log\left(\frac{p_n-\widetilde p_0}{\widetilde p_0} + 1\right).
$$
By Condition C3, we have that for $x\in[c_1,c_2]$ and $\balpha\in S_{\gamma}$,
\begin{eqnarray}
\left|\frac{p_n(x, \balpha) - \widetilde p_0(x, \balpha)}{\widetilde p_0(x, \balpha)}\right| &\lesssim & \left|\sum_{j=1}^M \alpha_{i,j} \{\cn_{h_j} \mathS_{h_j} f_{0,j}(x) - f_{0j}(x)\}\right| \nonumber \\
&\le & \sum_{j=1}^M \left|\cn_{h_j} \mathS_{h_j}f_{0,j}(x) - f_{0,j}(x)\right|. \label{var-2}
\end{eqnarray}
Applying Condition C3 again, we further note that
\begin{eqnarray}
\sup_{x\in [c_1, c_2]}|\cn_{h_j} \mathS_{h_j}f_{0,j}(x) - f_{0,j}(x)| &=& O(h). \label{var-3}
\end{eqnarray}
Hence
\begin{eqnarray}
\label{unform.order}
\sup_{x\in[c_1,c_2],~\balpha}\left|\frac{p_n(x, \balpha) - \widetilde p_0(x, \balpha)}{\widetilde p_0(x, \balpha)}\right|=O(h).
\end{eqnarray}
Applying the second-order Taylor expansion and using (\ref{unform.order}), we get that
\begin{eqnarray}
\log (p_n/\widetilde p_0)  =  \frac{p_n(x, \balpha) - \widetilde p_0(x, \balpha)}{\widetilde p_0(x, \balpha)} +R(x,\balpha),
\label{ratio.expand}
\end{eqnarray}
where the remaining term $R(x,\alpha)$ satisfies
\begin{eqnarray}
\label{unform.order2}
\sup_{x\in[c_1,c_2],~\balpha}|R(x,\balpha)|=O(h^2).
\end{eqnarray}

We now prove (\ref{bias-1}).
Combining (\ref{ratio.expand}) and (\ref{unform.order2}), we have that
\begin{eqnarray}
|E_0 \{ \log (p_n/\widetilde p_0) \}|&\leq& E_0\left| \frac{p_n(x, \balpha) - \widetilde p_0(x, \balpha)}{\widetilde p_0(x, \balpha)}\right| +O(h^2)\\
& \lesssim& \sum_{j=1}^M E_0 |\cn_{h_j} \mathS_{h_j}f_{0,j} - f_{0,j}|+O(h^2) \\
&=&O(h),
\end{eqnarray}
where we have used (\ref{var-2}) in the second step and (\ref{var-3})-(\ref{unform.order2}) in the third step.

Last, we show (\ref{var-1}).
Note that
\begin{eqnarray}
\var\left(\log\frac{p_n(X, \balpha)}{\widetilde p_0(X, \balpha)}\right) \le E_0 \log^2\left\{\frac{p_n(X, \balpha)}{\widetilde p_0(X, \balpha)}\right\}. \label{var-1-1}
\end{eqnarray}
 Combining (\ref{unform.order})--(\ref{unform.order2}) and (\ref{var-1-1}), we further get that
 \begin{eqnarray}
\var\left(\log\frac{p_n(X, \balpha)}{\widetilde p_0(X, \balpha)}\right) \le E_0\left[
\left\{\frac{p_n(x, \balpha) - \widetilde p_0(x, \balpha)}{\widetilde p_0(x, \balpha)}\right\}^2
\right]+O(h^3) = O(h^2). \label{var-1-2}
\end{eqnarray}
\qed

We finished verifying Conditions C1-C3 in Lemma \ref{lemma0.vm}.
Recall that
$$
\phi_n(\delta) = \delta^{1-1/(2a)}\sum_{j=1}^M (\log h_j)^{0.5}h_j^{-0.5-0.25/a}\left(1+\frac{1}{\sqrt{n} \delta^{1+1/(2a)}}\sum_{j=1}^M (\log h_j)^{0.5}h_j^{-0.5-0.25/a}\right)
$$
and
$
\MM_n(\widehat p, \widetilde p_0) \ge \MM_n(\widetilde p_0, \widetilde p_0) + O_p(h).
$
By applying Lemma \ref{lemma0.vm},  we have $d(\gamma \widehat p, \gamma \widetilde p_0) = O_p(r_n^{-1})$ with $r_n$ satisfying
\begin{eqnarray*}
r_n^2 \phi_n(1/r_n) &\le& \sqrt{n}\mbox{ and } r_n^{-2}=O_p(h)
\end{eqnarray*}
for every $a>0$.
Note that $r_n^2 \phi_n(1/r_n)\leq \sqrt{n}$ is equivalent to
$$
r_n^2 \delta^{1-1/(2a)}\sum_{j=1}^M (\log h_j)^{0.5}h_j^{-0.5-0.25/a} \lesssim \sqrt{n},
$$
which  implies that
$$r_n \lesssim \left(n^{0.5} \sum_{j=1}^M (\log h_j)^{0.5}h_j^{0.5+0.25/a} \right)^{1/(1+1/(2a))}. $$
For any $\vartheta > 0$, set $a$ sufficiently large, and $r_n^{-1} = O(h^{0.5}) + O(n^{-0.5 + \vartheta}h^{-0.5})$ we have
\begin{eqnarray*}
d(\gamma \widehat p, \gamma \widetilde p_0) = O_p(h^{0.5}) +O_p(n^{-0.5 + \vartheta}h^{-0.5}), \label{converge-d-1}
\end{eqnarray*}
which completes the proof of this theorem. \epf

\subsection*{Proof of Theorem \ref{theorem-3}}
In this subsection, we mainly establish the consistency of $\int_{\RR} |\cn_{h_j}\widehat f_j(x)-f_{0,j}(x)|dx$ as claimed in Theorem \ref{theorem-3} by using the consistency result for $d(\gamma \widehat p, \gamma \widetilde p_0)$ in Theorem \ref{consistency-theorem-1}.
We need the following lemma.

\begin{lemma}\label{proposition-3}
Assume Condition 4. For any $p(x, \balpha) = \sum_{j=1}^M \alpha_j \cn_{h_j} f_j(x) \in \mathP_n$, we have
\begin{eqnarray*}
\int_{\RR} |\cn_{h_j}f_j(x)-f_{0,j}(x)| dx \lesssim d(\gamma p, \gamma \widetilde p_0).
\end{eqnarray*}
\end{lemma}

\pf With  $\mathO_j, j = 1, \ldots, M$ and $\balpha_{0,j}$ given in Condition 4, we have
\begin{eqnarray}
&&\sum_{j=1}^M \int_{\RR} \left\{\sqrt{p(x, \balpha_{0,j})} - \sqrt{\widetilde p_0(x, \balpha_{0,j})}\right\}^2 dx \nonumber\\ &\lesssim& \sum_{j=1}^M \int\int _{\balpha\in \mathO_j; x \in \RR} \left(\sqrt{p(x,\balpha)} - \sqrt{\widetilde p_0(x,\balpha)}\right)^2 \gamma(\balpha) dx d\balpha \nonumber\\
&\le& d^2(\gamma p, \gamma \widetilde  p_0), \nonumber
\end{eqnarray}
which indicates for every $j = 1,\ldots, M$,
\begin{eqnarray}
\int_{\RR} \left\{\sqrt{p(x, \balpha_{0,j})} - \sqrt{\widetilde p_0(x, \balpha_{0,j})}\right\}^2 dx \lesssim d^2(\gamma p, \gamma \widetilde p_0). \label{prop-2}
\end{eqnarray}

Next we show that $\int_{\RR} |\cn_{h_j}f_j(x)-f_{0,j}(x)|dx$ can be bounded by a linear combination of the left hand side of (\ref{prop-2}). We need some notations.
Let
$A = (\balpha_{0,1}, \ldots, \balpha_{0,M})$ be an $M\times M$ invertible matrix
and denote
 $$A^{-1} = \big(a_{j,k}\big)_{{j=1,\ldots,M;~k = 1,\ldots,M}}.$$
Then
\begin{eqnarray}
\cn_{h_j} f_j (x)= \sum_{k=1}^M a_{j,k} p(x, \balpha_{0,k}),~~
f_{0,j}(x)=\sum_{k=1}^M a_{j,k} \widetilde p_0(x, \balpha_{0,k}). \nonumber
\end{eqnarray}
Therefore,
\begin{eqnarray}
&&\int_{\RR}|\cn_{h_j} f_j (x) - f_{0,j}(x)| dx \nonumber \\
& \le& \sum_{j=1}^M |a_{j,k}|\int_{\RR}|p(x, \balpha_{0,k}) - \widetilde p_0(x, \balpha_{0,k})| dx \nonumber \\
&\le& \sum_{j=1}^M |a_{j,k}| \sqrt{\int_{\RR} \left\{\sqrt{p(x, \balpha_{0,j})} - \sqrt{\widetilde p_0(x, \balpha_{0,j})}\right\}^2 dx} \cdot \sqrt{\int_{\RR} \left\{\sqrt{p(x, \balpha_{0,j})} +\sqrt{\widetilde p_0(x, \balpha_{0,j})}\right\}^2 dx} \nonumber \\ \label{eq-cnf-con-1}\\
&\le& \sum_{j=1}^M |a_{j,k}| d(\gamma p, \gamma \widetilde p_0)\sqrt{2\int_{\RR} \left\{ p(x, \balpha_{0,j}) + \widetilde p_0(x, \balpha_{0,j})\right\} dx} \label{eq-cnf-con-2}\\
&\lesssim& d(\gamma p, \gamma \widetilde p_0), \label{eq-cnf-con-3}
\end{eqnarray}
where from (\ref{eq-cnf-con-1}) to (\ref{eq-cnf-con-2}), we use (\ref{prop-2}) and the fact that $(a+b)^2 \le 2(a^2+b^2)$; from (\ref{eq-cnf-con-2}) to (\ref{eq-cnf-con-3}), we have applied Lemma \ref{proposition-1}, specifically,
\begin{eqnarray*}
\int_{\RR}p(x, \balpha_{0,j}) = \sum_{j=1}^M\alpha_{0,j} \int_{\RR}\cn_{h_j} f_j(x)dx \le \sum_{j=1}^M \alpha_{0,j} =  1,
\end{eqnarray*}
and likewise $\int_{\RR}\widetilde p_0(x, \balpha_{0,j}) \le  1$.
\epf

Combining Theorem \ref{consistency-theorem-1} and Lemma
\ref{proposition-3},  we immediately conclude the consistency of \\$\int_{\RR}|\cn_{h_j} \widehat f_j(x)-f_{0,j}(x)| dx$. That is: for any $\vartheta>0$,  we have
\begin{eqnarray}
\int_{\RR}|\cn_{h_j} \widehat f_j(x)- f_{0,j}(x)| dx= O_p(h^{0.5}) + O_p(n^{-0.5+\vartheta}h^{-0.5}), \label{con-thy-2-1}
\end{eqnarray}
which completes our proof of Theorem \ref{theorem-3}.

\subsection*{Proof of Theorem \ref{theorem-4}}
In this subsection, we prove Theorem \ref{theorem-4}, which  establishes  the $L_1$ consistency of $\widehat f_j(x)$, $j=1,\ldots,M$. Recall that
\begin{equation}
\label{appendix.hatf}
\widehat f_j(x)=\frac{\sum_{i=1}^n \widehat w_{i,j} K_{h_j}(x-X_i)}{\sum_{i=1}^n\widehat w_{i,j} }
\end{equation}
with
\begin{eqnarray*}
\widehat w_{i,j} &=& \frac{\alpha_{i,j}\cn_{h_j} \widehat f_j(X_i)}{\widehat p(X_i, \balpha_i)} \\
\widehat p(y, \balpha) &=& \sum_{s=1}^M \alpha_s \cn_{h_s} \widehat f_s(y).
\end{eqnarray*}
We investigate the asymptotic properties of the numerator and denominator of (\ref{appendix.hatf}) separately, and then establish the consistency of $\widehat f_j(x)$.  Based on Condition 3, we can find a constant $c>0$, such that $\inf_{x, \balpha}\widetilde p_0(x, \balpha) >2c$.  With straightforward manipulation, we note that $\widehat f_j(x)$ given in (\ref{appendix.hatf}) can be decompose as follows.
\begin{eqnarray}
\widehat f_j(x) = \frac{I_1(x)}{I_2} = \frac{I_{1,1}(x) - I_{1,2}(x) + I_{1,3}(x)}{I_2}, \label{hat-f-j-proof}
\end{eqnarray}
where
\begin{eqnarray}
I_{1,1}(x) &=& \frac{1}{n}\sum_{i=1}^n K_{h_j}(x-X_i)\frac{\alpha_{i,j}\cn_{h_j} \widehat f_j(X_i)}{\widehat p(X_i, \balpha_i)}I\{\widehat p(X_i, \alpha_i)\le c\} \label{I-1-1-def}\\
I_{1,2}(x) &=& \frac{1}{n}\sum_{i=1}^n K_{h_j}(x-X_i)\frac{\alpha_{i,j}\cn_{h_j} \widehat f_j(X_i)}{c}I\{\widehat p(X_i, \balpha_i)\le c\} \label{I-1-2-def}\\
I_{1,3}(x) &=& \frac{1}{n}\sum_{i=1}^n K_{h_j}(x-X_i)\frac{\alpha_{i,j}\cn_{h_j} \widehat f_j(X_i)}{\widehat p(X_i, \balpha_i)}I\{\widehat p(X_i, \balpha_i) > c\} \nonumber\\
&&+ \frac{1}{n}\sum_{i=1}^n K_{h_j}(x-X_i)\frac{\alpha_{i,j}\cn_{h_j} \widehat f_j(X_i)}{c}I\{\widehat p(X_i, \balpha_i) \le c\} \nonumber\\
&=& \frac{1}{n}\sum_{i=1}^n K_{h_j}(x-X_i)\frac{\alpha_{i,j}\cn_{h_j} \widehat f_j(X_i)}{\widehat p(X_i, \balpha_i)I\{\widehat p(X_i, \balpha_i) > c\} + c I\{\widehat p(X_i, \balpha_i) \le c\}} \label{I-1-3-def}\\
I_2 &=& \frac{1}{n} \sum_{i=1}^n \widehat w_{i,j}. \label{I-2-def}
\end{eqnarray}
Next we study the asymptotic behaviors of $I_{1,1}(x), I_{1,2}(x)$, and $I_{1,3}(x)$.  Studying $I_2$ is very similar but easier.  We first consider $I_{1,3}(x)$. We can write
\begin{eqnarray}
I_{1,3}(x) &=& \PP \left\{ K_{h_j}\left(Y-x\right) \cdot \{g_{j,c}(Y,\balpha) - g_{j,0}(Y, \balpha)\}\right\}  + \PP \left\{ K_{h_j}\left(Y-x\right) \cdot g_{j,0}(Y, \balpha)\right\}, \nonumber \\
&=& I_{1,3,1}(x) + I_{1,3,2}(x),\label{I-1-3}
\end{eqnarray}
where $``\PP"$ is operated on $(Y, \balpha)$; $g_{j,0}(y, \balpha) = \frac{\alpha_j f_{0,j}(y)}{\widetilde p_0(y, \balpha)}$; $g_{j,c}(y, \balpha) = \frac{\alpha_j\cn_{h_j} \widehat f_j(y)}{\widehat p(y, \balpha)I\{\widehat p(y, \balpha) > c\} + c I\{\widehat p(y, \balpha) \le c\}}$. We shall work on the following function classes.
\begin{itemize}
\item $K_{h_j}(y-x) \in \mathF_{K,j} = \left\{K_{h_j}(y-x): \mbox{indexed by } x \right\}$.
\item Recalling $\mathP_{n,j}$ defined in the proof of Lemma \ref{entropy}, we have $$\hspace{-0.2 in}g_{j,c}(y,\balpha)\in \mathF_{c,j} = \left\{\frac{\alpha_jf_j(y)}{p(y,\balpha)I\{p(y,\balpha) > c\} + c I\{p(y, \balpha) \le c\}}: p(y, \balpha) = \sum_{s=1}^M \alpha_s f_s(y), f_s\in \mathP_{n,s}\right\}.$$
\item We also need $|g_{j,c}(y,\balpha)-g_{j,0}(x,\balpha)|\in \widetilde \mathF_{c,j} = \{|g_{j,c} - g_{j,0}|: g_{j,c} \in \mathF_{c,j}\}$.
\end{itemize}

The following lemma calculates the bracketing numbers of the function classes given above.

\begin{lemma}\label{bracketing-numbers}
The bracketing numbers for $\mathF_{K,j}$, $\mathF_{c,j}$, and $\widetilde \mathF_{c,j}$ are given below. For every $\epsilon>0$,
\begin{itemize}
\item[(P1).] $N_{[]}(\epsilon, \mathF_{K,j}, L_2(P_0))\lesssim \frac{1}{h_j^2 \epsilon}$;
\item[(P2).] for an arbitrary $a>0$, $\log N_{[]}(\epsilon, \mathF_{c,j}, L_2(P_0)) \lesssim  \frac{\log h_j}{\epsilon^{1/a} h_j^{1+1/a}}$;
\item[(P3).] for an arbitrary $a>0$, $\log N_{[]}\left(\epsilon, \widetilde \mathF_{c,j}, L_2(P_0)\right) \lesssim  \frac{\log h_j}{\epsilon^{1/a} h_j^{1+1/a}}$.
\end{itemize}
In above, ``$\lesssim$" are up to universal constants depending on the upper bound of $K(\cdot)$, $a$, $c$, and $M$.

\end{lemma}

\pf Applying Theorem 2.7.11 in VM, (P1) immediately follows. We proceed to show (P2). Using an exactly the same strategy as the proof of (\ref{max-cond-3-2}) in Lemma \ref{entropy}, we can verify
\begin{eqnarray*}
\log N_{[]}(\epsilon, \mathP_{n,j}, L_2(P_0))\lesssim \frac{\log h_j}{\epsilon^{1/a} h_j^{1+1/a}}.
\end{eqnarray*}
For notational convenience, we write $N_j = N_{[]}(\epsilon, \mathP_{n,j}, L_2(P_0))$. Then for every $j$, there exist a set of $\epsilon$-brackets $\mathB_j = \{[u_{i,j}, v_{i,j}]: i = 1,\ldots, N_j\}$ that covers $\mathP_{n,j}$. We consider
\begin{eqnarray*}
\widetilde \mathB_j =\left\{[g_L(y, \balpha), g_U(y, \balpha)]: \begin{array}{l}g_L(y,\balpha) = \frac{\alpha_j u_{i_j,j}}{p_{U}}; \ \ g_U(y,\balpha) = \frac{\alpha_j v_{i_j,j}}{p_{L}};  \\   p_{U} = \widetilde p_{U}I\{\widetilde p_{U} > c\} + c I\{\widetilde p_{U} \le c\}; \ \ \widetilde p_{U} =  \sum_{l=1}^M \alpha_{l} v_{i_l, l};\\
  p_{L} = \widetilde p_{L}I\{\widetilde p_{L} > c\} + c I\{\widetilde p_{L} \le c\}; \ \ \widetilde p_{L} =  \sum_{l=1}^M \alpha_{l} u_{i_l, l}; \\
\mbox{for every } i_l = 1,\ldots,N_j; \quad \mbox{and} \quad  l=1,\ldots,M \end{array} \right\},
\end{eqnarray*}
which contains $\Pi_{j=1}^M N_j$ number of brackets.
We verify that $\widetilde \mathB_j$ covers $\mathF_{c,j}$. In fact, for every
\begin{eqnarray}
g_{j,c}(y, \balpha) =  \frac{\alpha_jf_j(y)}{p(y,\balpha)I\{p(y,\balpha) > c\} + c I\{p(y, \balpha) \le c\}} \in \mathF_{c,j}, \label{g-j-c}
\end{eqnarray}
since for every $j =1,\ldots,M$, $\mathB_j$ covers $\mathP_{n,j}$, there exist $(i_1, \ldots, i_M)$, where $1\le i_l \le N_l$ for every $l=1,\ldots, M$, such that
\begin{itemize}
\item[(C1).]
$\alpha_j u_{i_j,j} \le \alpha_j f_j \le \alpha_j v_{i_j,j}$; and

\item[(C2).] $\widetilde p_L \le p \le \widetilde p_U$, where $\widetilde p_L = \sum_{l=1}^M \alpha_l u_{i_l, l}$ and $\widetilde p_U = \sum_{l=1}^M \alpha_l v_{i_l, l}$.

\end{itemize}
Furthermore, note the fact that for any two functions $g_1$ and $g_2$, $g_1 \le g_2$ implies $g_1I\{g_1>c\} + cI\{g_1\le c\} \le g_2I\{g_2>c\} + cI\{g_2\le c\}$, where $c>0$ is an arbitrary constant. This together with $(C2)$ above leads to
\begin{itemize}
\item[(C3).] $p_L \le p I\{p>c\} + cI\{p\le c\} \le p_U$, where $p_L = \widetilde p_L I\{\widetilde p_L >c\} + cI\{\widetilde p_L >c\}$ and $p_U = \widetilde p_U I\{\widetilde p_U>c\} + cI\{\widetilde p_U \le c\}$.
\end{itemize}
(C1) and (C3) imply $g_L\le g_{j,c}\le g_U$, where $g_L = \frac{\alpha_j u_{i_j,j}}{p_{U}}$, $g_U = \frac{\alpha_j v_{i_j,j}}{p_{L}}$; $[g_L, g_U]$ is a bracket in $\widetilde \mathB_j$. Therefore, we have verified that $\widetilde \mathB_j$ covers $\mathF_{c,j}$.

We need to calculate the sizes of the brackets in $\widetilde \mathB_j$ under $L_2(P_0)$. To this end, we consider an arbitrary $[g_L, g_U] \in \widetilde \mathB_j$. Noting the facts that $|p_U-p_L|\le |\widetilde p_U-\widetilde p_L|$, $0\le \alpha_j \le 1$,  $0\le \alpha_j u_{i_j,j} \le p_L$ and $p_U \ge p_L \ge c >0$, we have
\begin{eqnarray*}
|g_U-g_L| &\le& \frac{\alpha_j}{p_L} \Big|v_{i_j,j} - u_{i_j,j}\Big| + \frac{\alpha_j u_{i_j,j}}{p_U p_L}\Big|p_U-p_L\Big| \\
&\le& \frac{|v_{i_j,j} - u_{i_j,j}|}{c} + \frac{|p_U-p_L|}{c} \\
&\le& \frac{|v_{i_j,j} - u_{i_j,j}|}{c} + \frac{|\widetilde p_U-\widetilde p_L|}{c} \\
&\le & \frac{|v_{i_j,j} - u_{i_j,j}|}{c} + \frac{1}{c}\sum_{l=1}^M |v_{i_l, l} - u_{i_l, l}|,
\end{eqnarray*}
which immediately leads to
\begin{eqnarray*}
&&\int_{x\in \RR} \int_{\balpha \in S_\gamma} |g_U(x, \balpha) - g_L(x, \balpha)|^2 \gamma(\balpha) \widetilde p_0(x, \balpha) dx d\balpha \\
&\lesssim & \sum_{l=1}^M \int_{x\in \RR} \int_{\balpha \in S_\gamma} |u_{i_l, l} - v_{i_l,l}|^2 \gamma(\balpha) \widetilde p_0(x, \balpha) dx d\balpha \\
&\lesssim & \epsilon^2,
\end{eqnarray*}
where the last ``$\lesssim$" is because that for every $l=1,\ldots,M$, $[u_{i_l, l}, v_{i_l,l}]$ is a $\epsilon$-bracket in $\mathB_j$ under $L_2(P_0)$. This together with the facts that $\widetilde \mathB_j$ covers $\mathF_{c,j}$ and $\widetilde \mathB_j$ contains $\Pi_{j=1}^M N_j$ number of brackets completes our proof for (P2) in this Lemma.

Last, we show (P3). Let $\mathF_{c,j,0} = \{g_{j,c} - g_{j,0}: g_{j,c} \in \mathF_{c,j}\}$. It is straightforward to check that
\begin{eqnarray}
\log N_{[]}\left(\epsilon, \mathF_{c,j, 0}, L_2(P_0)\right) \lesssim  \frac{\log h_j}{\epsilon^{1/a} h_j^{1+1/a}}. \label{entropy-mathF-c-j-0}
\end{eqnarray}
On the other hand, let $|f|$ be an arbitrary function in $\widetilde \mathF_{c,j}$ and $f \in \mathF_{c,j,0}$.  Let $[g_L, g_U]$ be the $\epsilon$-bracket in $\mathF_{c,j,0}$ such that $g_L \le f \le g_U$. By noting the fact that or any $y$ and $\balpha$, we must have either $g_L^+(y,\balpha) = 0$ or $g_U^-(y, \balpha) = 0$, we can easily check that
\begin{eqnarray}
g_L^+ + g_U^- \le |f|\le \max\{|g_L|, |g_U|\}, \label{bracket-1}
\end{eqnarray}
where for any function $g$, $g^- = -\min\{0, g\}$ and $g^+ = \max\{0, g\}$. Clearly $|g| = g^- + g^+$. Consequently,
\begin{eqnarray}
\left|\max\{|g_L|, |g_U|\} - g_L^+ - g_U^-\right| &\le& |g_L^++g_L^- - g_L^+ - g_U^-| + |g_U^++g_U^- - g_L^+ - g_U^-| \nonumber \\
&=& |g_L^- - g_U^-| + |g_U^+ - g_L^+| \le 2|g_U- g_L|. \label{bracket-2}
\end{eqnarray}
(\ref{bracket-1}) and (\ref{bracket-2}) imply that every $\epsilon$-bracket under $L_2(P_0)$ in $\mathF_{c,j,0}$ leads to a $2\epsilon$-bracket under $L_2(P_0)$ in $\widetilde \mathF_{c,j}$. This together with (\ref{entropy-mathF-c-j-0}) completes our proof of (P3) in this lemma.
\epf

With the lemma above, we study the asymptotic properties for $I_{1,3}$ given in (\ref{I-1-3}). We will consider $I_{1,3,1}(x)$ and $I_{1,3,2}(x)$ separately. First, we show
\begin{eqnarray}
\int_{\RR} |I_{1,3,1}(x)|dx = O_p\left(\frac{\sqrt{\log h_j}}{n^{0.5}h_j^{0.5+0.5/a}}\right) + d(\gamma \widehat p, \gamma\widetilde p_0). \label{I-1-3-1-proof-1}
\end{eqnarray}
To this end, note that
\begin{eqnarray}
\int_{\RR} |I_{1,3,1}(x)|dx &\le& \PP \left\{\int_{\RR} K_{h_j}(y-x) dx \cdot |g_{j,c}(y,\balpha)-g_{j,0}(y, \balpha)|\right\} \nonumber \\
&=& \PP \left\{|g_{j,c}(y,\balpha)-g_{j,0}(y, \balpha)|\right\} \label{I-1-3-1-proof-2}
\end{eqnarray}
where $\PP$ is operated on $y$ and $\alpha$. Now that $|g_{j,c}(y,\balpha)-g_{j,0}(y, \balpha)| \in \widetilde \mathF_{c,j}$, and for any function $f\in \widetilde \mathF_{c,j}$, we have
\begin{eqnarray*}
P_0 f^2 &\le& 4\\
\|f\|_\infty &\le& 2,
\end{eqnarray*}
which incorporated with Lemma 3.4.2 in VM lead to
\begin{eqnarray}
E_{P_0}\|\GG\|_{\widetilde \mathF_{c,j}} \lesssim \widetilde J_{[]}\left(2, \widetilde \mathF_{c,j}, L_2(P_0)\right)\left\{1+ \frac{\widetilde J_{[]}\left(2, \widetilde \mathF_{c,j}, L_2(P_0)\right)}{\sqrt{n}\cdot 2}\cdot4\right\}. \label{widetilde-mathF-c-j-1}
\end{eqnarray}
On the other hand, by (P3) in Lemma \ref{bracketing-numbers}, we have
\begin{eqnarray*}
\widetilde J_{[]}\left(2, \widetilde \mathF_{c,j}, L_2(P_0)\right) &\lesssim& \int_0^2 \sqrt{1+\frac{\log h_j}{\epsilon^{1/a} h_j^{1+1/a}}} \\
&\lesssim& \frac{\sqrt{\log h_j}}{h_j^{0.5+0.5/a}},
\end{eqnarray*}
which together with (\ref{widetilde-mathF-c-j-1}) leads to
\begin{eqnarray}
E_{P_0}\|\GG\|_{\widetilde \mathF_{c,j}} \lesssim \frac{\sqrt{\log h_j}}{h_j^{0.5+0.5/a}}. \label{widetilde-mathF-c-j-2}
\end{eqnarray}
By Chebyshev's inequality, (\ref{widetilde-mathF-c-j-2}) immediately implies
\begin{eqnarray}
\PP \left\{|g_{j,c}(y,\balpha)-g_{j,0}(y, \balpha)|\right\} - P_0\left\{|g_{j,c}(y,\balpha)-g_{j,0}(y, \balpha)|\right\} = \frac{\sqrt{\log h_j}}{n^{0.5}h_j^{0.5+0.5/a}}, \label{I-1-3-1-proof-3}
\end{eqnarray}
where the convergence of $P_0\left\{|g_{j,c}(y,\balpha)-g_{j,0}(y, \balpha)|\right\}$ is studied by the following lemma.

\begin{lemma} \label{lemma-I-1-3-1-proof-1}
Recall $g_{j,0}(y, \balpha) = \frac{\alpha_j f_{0,j}(y)}{\widetilde p_0(y, \balpha)}$; $g_{j,c}(y, \balpha) = \frac{\alpha_j\cn_{h_j} \widehat f_j(y)}{\widehat p(y, \balpha)I\{\widehat p(y, \balpha) > c\} + c I\{\widehat p(y, \balpha) \le c\}}$. We have
\begin{eqnarray}
P_0\left\{|g_{j,c}(y,\balpha)-g_{j,0}(y, \balpha)|\right\} \lesssim d(\gamma\widehat p, \gamma \widetilde p_0). \label{P-0-g-c-g-0}
\end{eqnarray}
\end{lemma}

\pf Note the fact
\begin{eqnarray*}
g_{j,c}(y, \balpha) &=& \frac{\alpha_j\cn_{h_j} \widehat f_j(y)}{\widehat p(y, \balpha)I\{\widehat p(y, \balpha) > c\} + c I\{\widehat p(y, \balpha) \le c\}} \\
&=& \frac{\alpha_j\cn_{h_j} \widehat f_j(y)}{\widehat p(y, \balpha)}I\{\widehat p(y, \balpha) > c\} + \frac{\alpha_j\cn_{h_j} \widehat f_j(y)}{c}I\{\widehat p(y, \balpha) \le c\}.
\end{eqnarray*}
Therefore
\begin{eqnarray}
P_0\left\{|g_{j,c}(Y,\balpha)-g_{j,0}(Y, \balpha)|\right\} \le I_{1,3,1,1} + I_{1,3,1,2} + I_{1,3,1,3}, \label{P-0-g-c-g-0-1}
\end{eqnarray}
where
\begin{eqnarray*}
I_{1,3,1,1} &=& \int_{\RR} \int_{\balpha \in S_\gamma} \left|\frac{\alpha_j\cn_{h_j} \widehat f_j(y)}{\widehat p(y, \balpha)}-\frac{\alpha_j f_{0,j}(y)}{\widetilde p_0(y, \balpha)}\right| I\{\widehat p(y, \balpha) > c\} \gamma(\balpha) \widetilde p_0(y,\balpha) d\balpha dy \\
I_{1,3,1,2} &=& \int_{\RR} \int_{\balpha \in S_\gamma} \frac{\alpha_j\cn_{h_j} \widehat f_j(y)}{c}I\{\widehat p(y, \balpha) \le c\}\gamma(\balpha) \widetilde p_0(y,\balpha) d\balpha dy\\
I_{1,3,1,3} &=& \int_{\RR} \int_{\balpha \in S_\gamma} \frac{\alpha_j f_{0,j}(y)}{\widetilde p_0(y, \balpha)} I\{\widehat p(y, \balpha) \le c\} \gamma(\balpha) \widetilde p_0(y,\balpha) d\balpha dy.
\end{eqnarray*}
To show this lemma, we only need to bound $I_{1,3,1,1}$, $I_{1,3,1,2}$ and $I_{1,3,1,3}$. We first consider $I_{1,3,1,1}$:
\begin{eqnarray}
I_{1,3,1,1} &\le& \int_{\RR} \int_{\balpha \in S_\gamma} \frac{\alpha_j }{\widehat p(y, \balpha)} \left|\cn_{h_j}\widehat f_j(y) - f_{0,j}(y)\right|I\{\widehat p(y, \balpha) > c\}\gamma(\balpha)\widetilde p_0(y, \balpha) d\balpha dy \nonumber \\
&& + \int_{\RR} \int_{\balpha \in S_\gamma} \frac{\alpha_j f_{0,j}(y) }{\widehat p(y, \balpha) \widetilde p_0(y,\balpha)} \left|\widehat p(y, \balpha) - \widetilde p_0(y, \balpha)\right|I\{\widehat p(y, \balpha) > c\}\gamma(\balpha)\widetilde p_0(y, \balpha)  d\balpha dy \nonumber\\
&\lesssim& \int_{\RR} \left| \cn_{h_j}\widehat f_j(y) - f_{0,j}(y) \right| d\balpha dy + \int_{\RR} |\widehat p(y, \balpha)- \widetilde p_0(y, \balpha)|\gamma(\balpha) d\balpha dy \nonumber\\
&\lesssim& \int_{\RR} \left| \cn_{h_j}\widehat f_j(y) - f_{0,j}(y) \right|  d\balpha dy + d(\gamma \widehat p, \gamma \widetilde p_0) \nonumber\\
&\lesssim&d(\gamma \widehat p, \gamma \widetilde p_0), \label{I-1-3-1-1-0}
\end{eqnarray}
where for the last ``$\lesssim$", we have applied Lemma \ref{proposition-3}. Next, we consider $I_{1,3,1,2}$ and $I_{1,3,1,3}$ together, it is clearly seen that
\begin{eqnarray}
&&I_{1,3,1,2} \lesssim I_{1,3,1,4}, \nonumber \\
\mbox{and} \quad && I_{1,3,1,3} \lesssim I_{1,3,1,4}, \label{I-1-3-1-4-0}
\end{eqnarray}
where
\begin{eqnarray*}
I_{1,3,1,4} = \int_{\RR} \int_{\balpha \in S_\gamma} I\{\widehat p(y, \balpha)  \le c\} \gamma(\balpha)\widetilde p_0(y, \balpha)d\balpha dy.
\end{eqnarray*}
Recalling the definition of $c$: $\inf_{x, \balpha}\widetilde p_0(x, \balpha) >2c$, we have
\begin{eqnarray}
I_{1,3,1,4} &\le& \int_{\RR} \int_{\balpha \in S_\gamma} I\{|\widetilde p_0(x, \balpha) - \widehat p(y, \balpha)| > c\}\gamma(\balpha)\widetilde p_0(y, \balpha)d\balpha dy \nonumber\\
&\le& \int_{\RR} \int_{\balpha \in S_\gamma} I\{|\widetilde p_0(x, \balpha) - \widehat p(y, \balpha)| > c\} \frac{|\widetilde p_0(x, \balpha) - \widehat p(y, \balpha)| }{c}\gamma(\balpha)\widetilde p_0(y, \balpha)d\balpha dy \nonumber\\
&\lesssim&\int_{\RR} \int_{\balpha \in S_\gamma} |\widetilde p_0(x, \balpha) - \widehat p(y, \balpha)| \gamma(\balpha)\widetilde p_0(y, \balpha)d\balpha dy \nonumber\\
&\lesssim& d(\gamma \widehat p, \gamma \widetilde p_0). \label{I-1-3-1-4-1}
\end{eqnarray}
Combining (\ref{P-0-g-c-g-0-1}), (\ref{I-1-3-1-1-0}), (\ref{I-1-3-1-4-0}), and (\ref{I-1-3-1-4-1}), we immediately conclude (\ref{P-0-g-c-g-0}).
\epf

Combining Lemma \ref{lemma-I-1-3-1-proof-1}, (\ref{I-1-3-1-proof-2}), and (\ref{I-1-3-1-proof-3}) leads to (\ref{I-1-3-1-proof-1}). Second, we verify
\begin{eqnarray}
\sup_{x}\left|I_{1,3,2}(x)-\int_{\RR} K_{h_j}(y-x) f_{0,j}(y)dy \int_{\balpha\in S_\gamma} \alpha_j \gamma(\balpha) d\balpha \right| = O_p\left(\frac{1}{\sqrt{nh_j}}\right). \label{I-1-3-2-proof-1}
\end{eqnarray}
Recall the definition of $I_{1,3,2}(x)$ in (\ref{I-1-3}) and $g_{j,0}(y, \balpha) = \frac{\alpha_j f_{0,j}(y)}{\widetilde p_0(y, \balpha)}$. We have
\begin{eqnarray}
I_{1,3,2}(x) = \PP \left\{ K_{h_j}\left(Y-x\right) \cdot g_{j,0}(Y, \balpha)\right\} = \PP\left\{g_{x,j,0}(Y, \balpha)\right\} \label{I-1-3-2-proof-2}
\end{eqnarray}
where $g_{x,j,0}(y, \balpha) = K_{h_j}(y-x)g_{j,0}(y, \balpha)$. For every $n$, we consider the class of functions $\mathF_{x,j,0} =  \{g_{x,j,0}: \mbox{indexed by } x\}$.  Then, it is readily checked that for every $g_{x, j, 0} \in \mathF_{x,j,0}$
\begin{eqnarray*}
&&P_0 g_{x,j,0}^2(Y, \balpha) \lesssim 1/h_j \\
&\mbox{and}& \|g_{x,j,0}\|_{\infty} \lesssim 1/h_j,
\end{eqnarray*}
which incorporated with Lemma 3.4.2 in VM lead to
\begin{eqnarray}
E_{P_0}\|\GG_n\|_{\mathF_{x,j, 0}} \lesssim \widetilde J_{[]}(1/\sqrt{h_j}, \mathF_{x,j, 0}, L_2(P_0))\left\{1+ \frac{\widetilde J_{[]}(1/\sqrt{h_j}, \mathF_{x,j, 0}, L_2(P_0))}{\sqrt{n}/h_j}\frac{1}{h_j}\right\}. \label{thm-4-3}
\end{eqnarray}
On the other hand, we have $\mathF_{x, j, 0} = g_{j,0}(y, \balpha) \cdot \mathF_{K,j}$, with $\mathF_{K,j} = \{K_{h_j}(y-x): \mbox{indexed by } x\}$. Recall (P1) in Lemma \ref{bracketing-numbers}, for every $\epsilon >0$, we have
\begin{eqnarray*}
N_{[]}(\epsilon, \mathF_{K,j}, L_2(P_0)) \lesssim \frac{1}{h_j^2 \epsilon},
\end{eqnarray*}
which together with the fact that the single function $|g_{j,0}(y,\balpha)|\le 1$ implies
\begin{eqnarray*}
N_{[]}(\epsilon, \mathF_{x, j, 0}, L_2(P_0)) \lesssim \frac{1}{h_j^2 \epsilon},
\end{eqnarray*}
and therefore
\begin{eqnarray*}
\widetilde J_{[]}(1/\sqrt{h_j}, \mathF_{x,j, 0}, L_2(P_0)) \lesssim h_j^{-b} h_j^{-0.5(1-b/2)},
\end{eqnarray*}
for any arbitrary $0<b<1$. The above ``$\lesssim$" is up to a universal constant not depending on $b$. Setting $b\to 0+$, we have
\begin{eqnarray*}
\widetilde J_{[]}(1/\sqrt{h_j}, \mathF_{x,j, 0}, L_2(P_0)) \lesssim h_j^{-0.5},
\end{eqnarray*}
which together with (\ref{thm-4-3}) leads to
\begin{eqnarray}
E_{P_0}\|\GG_n\|_{\mathF_{x,j, 0}} \lesssim 1/\sqrt{h_j}. \label{thm-4-5}
\end{eqnarray}
By Chebyshev's inequality, (\ref{thm-4-5}) immediately implies
\begin{eqnarray}
\sup_{x}|\PP g_{x,j,0}(Y,\balpha) - P_0g_{x,j,0}(Y, \balpha)| = O_p(1/\sqrt{nh_j}).  \label{thm-4-6}
\end{eqnarray}
Furthermore, one can easily check
\begin{eqnarray*}
P_0g_{x,j,0}(Y, \balpha) = \int_{\RR} K_{h_j}(y-x) f_{0,j}(y)dy \int_{\balpha\in S_\gamma} \alpha_j \gamma(\balpha) d\balpha,
\end{eqnarray*}
which together with (\ref{I-1-3-2-proof-2}) and (\ref{thm-4-6}) leads to (\ref{I-1-3-2-proof-1}). Now we combine (\ref{I-1-3-1-proof-1}), (\ref{I-1-3-2-proof-1}), with (\ref{I-1-3}) and conclude
\begin{equation}
\int_{\RR}\left|I_{1,3}(x) - \int_{\RR} K_{h_j}(y-x) f_{0,j}(y)dy \int_{\balpha\in S_\gamma} \alpha_j \gamma(\balpha) d\balpha\right|dx = O_p\left(\frac{\sqrt{\log h_j}}{n^{0.5}h_j^{0.5+0.5/a}}\right) + d(\gamma \widehat p, \gamma \widetilde p_0). \label{I-1-3-proofed}
\end{equation}

We proceed to consider the consistency of $I_{1,1}(x)$ and $I_{1,2}(x)$. Note that they are respectively defined in (\ref{I-1-1-def}) and (\ref{I-1-2-def}). Recall
\begin{eqnarray}
I_{1,1}(x) &=& \frac{1}{n}\sum_{i=1}^n K_{h_j}(x-X_i)\frac{\alpha_{i,j}\cn_{h_j} \widehat f_j(X_i)}{\widehat p(X_i, \balpha_i)}I\{\widehat p(X_i, \alpha_i)\le c\} \le I_{1,4}(x)\nonumber\\
I_{1,2}(x) &=& \frac{1}{n}\sum_{i=1}^n K_{h_j}(x-X_i)\frac{\alpha_{i,j}\cn_{h_j} \widehat f_j(X_i)}{c}I\{\widehat p(X_i, \balpha_i)\le c\} \le I_{1,4}(x), \label{I-1-1-proof-1}
\end{eqnarray}
where
\begin{eqnarray*}
0\le I_{1,4}(x) = \frac{1}{n}\sum_{i=1}^n K_{h_j}(x-X_i)I\{\widehat p(X_i, \alpha_i)\le c\}.
\end{eqnarray*}
Recalling the definition of $c$: $\inf_{x, \balpha}\widetilde p_0(x, \balpha) >2c$, we have
\begin{eqnarray}
0&\le& \int_{\RR}I_{1,4}(x)dx \le  \frac{1}{n}\sum_{i=1}^n I\{\widehat p(X_i, \alpha_i)\le c\} \nonumber\\
&\le& \frac{1}{n} \sum_{i=1}^n I\{\widetilde p_0(X_i, \balpha) - \widehat p(X_i, \balpha)>c\} \nonumber\\
&\le& \frac{1}{nc} \sum_{i=1}^n I\{\widetilde p_0(X_i, \balpha) - \widehat p(X_i, \balpha)>c\}\left\{\widetilde p_0(X_i, \balpha) - \widehat p(X_i, \balpha)\right\} \nonumber\\
&\lesssim& \frac{1}{n} \sum_{i=1}^n I\{\widetilde p_0(X_i, \balpha) - \widehat p(X_i, \balpha) >0\}\left\{\widetilde p_0(X_i, \balpha) - \widehat p(X_i, \balpha)\right\} \nonumber\\
&=& \PP \left[ g_p(Y, \balpha) I\{g_p(Y, \balpha)>0\}\right], \label{I-1-1-proof-2}
\end{eqnarray}
where $g_p(y, \balpha) = \widetilde p_0(y, \balpha) - \widehat p(y, \balpha)$. Clearly $$ g_p(Y, \balpha) I\{g_p(Y, \balpha)>0\} \in \mathF_{p,I} = \{\left\{\widetilde p_0(y, \balpha) - p(y, \balpha)\right\}I\{\widetilde p_0(y, \balpha) - p(y, \balpha)>0\}: p\in \mathP_n\},$$ where we refer to (\ref{def-mathP-n}) for the definition of $\mathP_n$. In the lemma below, we establish the $\epsilon$-bracketing number of $\mathF_{p,I}$ under $L_2(P_0)$.

\begin{lemma} \label{bracketing-numbers-2}
For an arbitrary $\epsilon>0$ and $a>0$, we have
\begin{eqnarray*}
\log N_{[]}(\epsilon, \mathF_{p,I}, L_2(P_0)) \lesssim \frac{\log h_j}{\epsilon^{1/a} h_j^{1+1/a}}.
\end{eqnarray*}
\end{lemma}
\pf Using exactly the same procedure as Lemma \ref{entropy}, we have
\begin{eqnarray*}
\log N_{[]}(\epsilon, \mathP_n, L_2(P_0)) \lesssim \frac{\log h_j}{\epsilon^{1/a} h_j^{1+1/a}},
\end{eqnarray*}
which entails
\begin{eqnarray*}
\log N_{[]}(\epsilon, \mathP_{n,0}, L_2(P_0)) \lesssim \frac{\log h_j}{\epsilon^{1/a} h_j^{1+1/a}},
\end{eqnarray*}
where $\mathP_{n,0} = \widetilde p_0 - \mathP_0$.
On the other hand, let $[g_{L,i}, g_{U,i}]$, $i=1,\ldots, N_{[]}(\epsilon, \mathP_{n,0}, L_2(P_0))$ be the corresponding $\epsilon$-brackets for $\mathP_n$. We consider $[\widetilde g_{L,i}, \widetilde g_{U,i}]$, $i=1,\ldots,N_{[]}(\epsilon, \mathP_{n,0}, L_2(P_0))$, where $\widetilde g_{L,i} = g_{L,i}I\{g_{L,i}>0\}$ and $\widetilde g_{U,i} = g_{U,i}I\{g_{U,i}>0\}$. Since for any arbitrary functions $g_1$, $g_2$,  if $g_1\le g_2$, then $g_1I\{g_1>0\} \le g_2 I\{g_2>0\}$, we immediately conclude that the set of brackets $[\widetilde g_{L,i}, \widetilde g_{U,i}]$, $i=1,\ldots,N_{[]}(\epsilon, \mathP_{n,0}, L_2(P_0))$, covers $\mathF_{p,I}$. Furthermore, it is straightforward to check that
\begin{eqnarray*}
0\le \widetilde g_{U,i} - \widetilde g_{L,i} \le g_{U,i} - g_{L,i}.
\end{eqnarray*}
Therefore, we have
\begin{eqnarray*}
\log N_{[]}(\epsilon, \mathF_{p,I}, L_2(P_0)) \le \log N_{[]}(\epsilon, \mathP_{n,0}, L_2(P_0)) \lesssim \frac{\log h_j}{\epsilon^{1/a} h_j^{1+1/a}},
\end{eqnarray*}
which completes our proof of this lemma. \epf

We continue with our analysis of the asymptotic property for $I_{1,4}(x)$. Noting the fact that $\widetilde p_0$ is bounded, therefore $\mathF_{p,I}$ is uniformly bounded. For any function $f\in \mathF_{p,I}$, we have
\begin{eqnarray*}
P_0 f^2 \lesssim 1\\
\|f\|_\infty \lesssim 1,
\end{eqnarray*}
which incorporated with Lemma 3.4.2 in VM lead to
\begin{eqnarray}
E_{P_0}\|\GG_n\|_{\mathF_{p,I}} \lesssim \widetilde J_{[]}(1, \mathF_{p,I}, L_2(P_0))\left\{1+ \frac{\widetilde J_{[]}(1, \mathF_{p,I}, L_2(P_0))}{\sqrt{n}}\right\}.  \label{I-1-1-proof-3}
\end{eqnarray}
On the other hand, applying Lemma \ref{bracketing-numbers-2}, we have
\begin{eqnarray*}
\widetilde J_{[]}\left(1, \mathF_{p,I}, L_2(P_0)\right) &\lesssim& \int_0^1 \sqrt{1+\frac{\log h_j}{\epsilon^{1/a} h_j^{1+1/a}}} \\
&\lesssim& \frac{\sqrt{\log h_j}}{h_j^{0.5+0.5/a}},
\end{eqnarray*}
which together with (\ref{I-1-1-proof-3}) leads to
\begin{eqnarray}
E_{P_0}\|\GG_n\|_{\mathF_{p,I}} \lesssim \frac{\sqrt{\log h_j}}{h_j^{0.5+0.5/a}}. \label{I-1-1-proof-4}
\end{eqnarray}
By Chebyshev's inequality, (\ref{I-1-1-proof-4}) immediately implies
\begin{equation}
\PP \left[ g_p(Y, \balpha) I\{g_p(Y, \balpha)>0\}\right] - P_0\left[ g_p(Y, \balpha) I\{g_p(Y, \balpha)>0\}\right] = O_p\left(\frac{\sqrt{\log h_j}}{n^{0.5}h_j^{0.5+0.5/a}}\right). \label{I-1-1-proof-5}
\end{equation}
It is left to examine $P_0\left[ g_p(Y, \balpha) I\{g_p(Y, \balpha)>0\}\right]$. In fact
\begin{eqnarray}
&&P_0\left[ g_p(Y, \balpha) I\{g_p(Y, \balpha)>0\}\right] \nonumber \\
&=& \int_{\RR} \int_{\balpha \in S_\gamma}I\{\widetilde p_0(y, \balpha) - \widehat p(y, \balpha) >0\}\left\{\widetilde p_0(y, \balpha) - \widehat p(y, \balpha)\right\} \gamma(\balpha) \widetilde p_0(y, \balpha) d\balpha dy \nonumber \\
&\le& \int_{\RR} \int_{\balpha \in S_\gamma} |\widetilde p_0(y, \balpha) - \widehat p(y, \balpha)| \gamma(\balpha) \widetilde p_0(y, \balpha) d\balpha dy \nonumber\\
&\lesssim & d(\gamma \widehat p, \gamma \widetilde p_0). \label{I-1-1-proof-6}
\end{eqnarray}
Now, we combine (\ref{I-1-1-proof-1}), (\ref{I-1-1-proof-2}), (\ref{I-1-1-proof-5}), and (\ref{I-1-1-proof-6}) to conclude
\begin{eqnarray*}
\int_{\RR}I_{1,1}(x) dx \le \int_{\RR}I_{1,4}(x) dx\lesssim d(\gamma \widehat p, \gamma \widetilde p_0) + O_p\left(\frac{\sqrt{\log h_j}}{n^{0.5}h_j^{0.5+0.5/a}}\right)\\
\int_{\RR}I_{1,2}(x) dx \le \int_{\RR}I_{1,4}(x) dx \lesssim d(\gamma \widehat p, \gamma \widetilde p_0) + O_p\left(\frac{\sqrt{\log h_j}}{n^{0.5}h_j^{0.5+0.5/a}}\right),
\end{eqnarray*}
which together with (\ref{I-1-3-proofed}) and (\ref{hat-f-j-proof}) conclude
\begin{eqnarray}
\int_{\RR}\left|I_1(x) - \int_{\RR} K_{h_j}(y-x) f_{0,j}(y)dy \int_{\balpha\in S_\gamma} \alpha_j \gamma(\balpha) d\balpha\right| dx \lesssim d(\gamma \widehat p, \gamma \widetilde p_0) + O_p\left(\frac{\sqrt{\log h_j}}{n^{0.5}h_j^{0.5+0.5/a}}\right). \label{result-I-1}
\end{eqnarray}
With similar but easier procedures as above, we can verify
\begin{eqnarray}
\left|I_2 - \int_{\balpha\in S_\gamma} \alpha_j \gamma(\balpha) d\balpha\right| \lesssim d(\gamma \widehat p, \gamma \widetilde p_0) + O_p\left(\frac{\sqrt{\log h_j}}{n^{0.5}h_j^{0.5+0.5/a}}\right). \label{result-I-2}
\end{eqnarray}

We now prove Theorem \ref{theorem-4}.
Recall the definition of $I_1(x)$ and $I_2$ in (\ref{hat-f-j-proof}) and their asymptotic properties we have presented in (\ref{result-I-1}) and (\ref{result-I-2}). We have
\begin{eqnarray}
&&\int_{\RR} |\widehat f_j(x)-f_{0,j}(x)|dx  = \int_{\RR} |I_1(x)/I_2-f_{0,j}(x)|dx \nonumber\\
&\lesssim& \int_{\RR} |I_1(x)-I_2 f_{0,j}(x)| dx \nonumber\\
&\lesssim& \int_{\RR}\left|\int K_{h_j}(u-x) f_{0,j}(u) du - f_{0,j}(x)\right| dx \int \alpha_j \gamma (\balpha) d\balpha \nonumber \\
&& + d(\gamma \widehat p, \gamma \widetilde p_0) + O_p\left(\frac{\sqrt{\log h_j}}{n^{0.5}h_j^{0.5+0.5/a}}\right),
\end{eqnarray}
which together with Theorem \ref{consistency-theorem-1} and the following easily checked result (\ref{thm-4-12}) based on Conditions 2 and 3 completes our proof of this theorem by setting $a$ sufficiently large.
\begin{eqnarray}
\int_{\RR }\left|\int K_{h_j}(u-x) f_{0,j}(u) du - f_{0,j}(x)\right| dx &=& O(h). \endpf\label{thm-4-12}
\end{eqnarray}

\end{document}